\documentclass[aps,prb,twocolumn,showpacs]{revtex4}

\usepackage{graphicx}
\usepackage{amsmath}
\usepackage{amssymb}

\begin{document}

\title{\vspace{-10mm} Renormalization of electrons in bilayer cuprate superconductors}

\author{Yiqun Liu$^{1}$, Yu Lan$^{2}$, Yingping Mou$^{1}$, and
Shiping Feng$^{1}$\footnote{Corresponding author. E-mail address: spfeng@bnu.edu.cn}}


\affiliation{$^{1}$Department of Physics, Beijing Normal University, Beijing 100875, China}

\affiliation{$^{2}$College of Physics and Electronic Engineering, Hengyang Normal University,
Hengyang 421002, China}

\begin{abstract}
Cuprate superconductors have a layered structure, and then the physical properties are significantly
enriched when two or more two copper-oxide layers are contained in a unit cell. Here the characteristic
features of the renormalization of the electrons in the bilayer cuprate superconductors are investigated
within the framework of the kinetic-energy driven superconducting mechanism. It is shown that the
electron quasiparticle excitation spectrum is split into its bonding and antibonding components due to
the presence of the bilayer coupling, with each component that is independent. However, in the underdoped
and optimally doped regimes, although the bonding and antibonding electron Fermi surface contours deriving
from the bonding and antibonding layers are truncated to form the disconnected bonding and antibonding Fermi
arcs, almost all spectral weights in the bonding and antibonding Fermi arcs are reduced to the tips of the
bonding and antibonding Fermi arcs, which in this case coincide with the bonding and antibonding hot spots.
These hot spots connected by the scattering wave vectors ${\bf q}_{i} $ construct an {\it octet} scattering
model, and then the enhancement of the quasiparticle scattering processes with the scattering wave vectors
${\bf q}_{i}$ is confirmed via the result of the autocorrelation of the electron quasiparticle excitation
spectral intensities. Moreover, the peak-dip-hump structure developed in each component of the electron
quasiparticle excitation spectrum along the corresponding electron Fermi surface is directly related with
the peak structure in the quasiparticle scattering rate except for at around the hot spots, where the
peak-dip-hump structure is caused mainly by the pure bilayer coupling. Although the kink in the electron
quasiparticle dispersion is present all around the electron Fermi surface, when the momentum moves away
from the node to the antinode, the kink energy smoothly decreases, while the dispersion kink becomes more
pronounced, and in particular, near the cut close to the antinode, develops into a break separating of the
fasting dispersing high-energy part of the electron quasiparticle excitation spectrum from the slower
dispersing low-energy part. By comparing with the corresponding results in the single-layer case, the theory
also indicates that the characteristic features of the renormalization of the electrons are particularly
obvious due to the presence of the bilayer coupling.
\end{abstract}


\maketitle

\section{Introduction}\label{Introduction}

The parent compounds of cuprate superconductors are actually Mott insulators, which are known to be the
result from the very strong electron-electron correlation \cite{Anderson87}, superconductivity then is
tuned through a remarkable progression of the states of matter by doping charge carriers \cite{Bednorz86,Wu87}.
After over thirty years of extensive research, some issues in cuprtae superconductors have been settled,
such as the d-wave type electron pairing state \cite{Tsuei00}, which is one of the cornerstones of our
understanding of the unconventional mechanism of superconductivity in cuprate superconductors, whereas others
remain controversial, such as the nature of the electron quasiparticle excitation
\cite{Damascelli03,Campuzano04,Fink07,Kordyuk14,Zhou18,Carbotte11}, which plays an essential role in the
occurrence of superconductivity. In cuprate superconductors, the electrons interact strongly with various
bosonic excitations resulting in the modification to both their lifetime and binding-energy, a process
so-called renormalization of the electrons to form the electron quasiparticle excitations \cite{Damascelli03,Campuzano04,Fink07,Kordyuk14,Zhou18,Carbotte11}, and then all the anomalous properties
\cite{Damascelli03,Campuzano04,Fink07,Kordyuk14,Zhou18,Carbotte11,Kastner98,Timusk99,Hufner08,Comin16,Vishik18,Fischer07,Devereaux07}
arise from this renormalization of the electrons. In this case, the systematic study of the renormalization
of the electrons in cuprate superconductors allows the identification of the collective bosonic excitations
mediating the interaction between the electrons that is most likely also responsible for the exceptionally high
superconducting (SC) transition temperature $T_{\rm c}$.

The single common feature in the layered crystal structure of cuprate superconductors is the presence of the
two-dimensional copper-oxide layers \cite{Damascelli03,Campuzano04,Fink07,Kordyuk14,Zhou18,Carbotte11,Kastner98,Timusk99,Hufner08,Comin16,Vishik18,Fischer07,Devereaux07},
and then the unusual behaviors and the related superconductivity come from the strongly correlated motion of the
electrons in these copper-oxide layers. In particular, the strongly correlated motion of the electrons confined
to the copper-oxide layers has been confirmed experimentally by the incoherent charge-transport along the
out-of-copper-oxide layer direction \cite{Cooper94,Nakamura93,Hou94,Takenaka94}. However, the exotic properties
are significantly enriched by the coupling between the copper-oxide layers when two or more two copper-oxide
layers are contained in a unit cell
\cite{Vincini19,Kunisada17,Adachi15,Mukuda12,Johnston10,Iyo07,Eisaki04,Karppinen99,Stasio90,Tarascon88}.
For instance, $T_{\rm c}$ is very sensitive to the number of the copper-oxide layers $n$ within a unit cell \cite{Vincini19,Kunisada17,Adachi15,Mukuda12,Johnston10,Iyo07,Eisaki04,Karppinen99,Stasio90,Tarascon88}, i.e.,
for all families of cuprate superconductors, the magnitude of the optimized $T_{\rm c}$ is found experimentally
to increase with the increase of $n$ per unit cell for the case of $n<3$, and reaches the maximum for the case
of $n=3$, then decreases slightly and saturates for the case of $n>3$, which therefore shows clearly that the
number of the copper-oxide layers dependence of the optimized $T_{\rm c}$ is induced by the coupling between
the copper-oxide layers within a unit cell. However, up to now, the cause of this $T_{\rm c}$ enhancement is
still not clear. In this case, it is particularly significant to clarify the characteristic features of the
renormalization of the electrons in cuprate superconductors distinctive for the families in the presence of the
coupling between the copper-oxide layers within a unit cell.

In this paper, we focus on the coupling effect between the copper-oxide layers on the renormalization of the
electrons in cuprate superconductors. In particular, the bilayer cuprate superconductor is an ideal system to
tackle the coupling effect between the copper-oxide layers on the renormalization of the electrons, where the
bilayer splitting due to the presence of the bilayer interaction in a unit cell has been observed from the
angle-resolved photoemission spectroscopy (ARPES) experiments in a wide doping range
\cite{Feng-dl01,Chuang01,Eschrig06}, with the largest value of the bilayer splitting that appears at around
the antinodal region, and then decreases upon approaching to the nodal region. In particular, these ARPES
experimental observations also indicate that this bilayer splitting derives the electron quasiparticle
excitation spectrum into the bonding and antibonding components \cite{Feng-dl01,Chuang01,Eschrig06}, which
leads to form the bonding and antibonding electron Fermi surface (EFS) contours corresponding to different
doping levels with distinct SC gap \cite{Liu92,Kordyuk04,Chuang04,Borisenko04,Loret18,Ai19,Gao-Q19}. Moreover,
it has been shown that this bilayer coupling may play a main role in the form of the well-known peak-dip-hump
(PDH) structure in the electron quasiparticle excitation spectrum of the bilayer cuprate superconductors in
the overdoped regime \cite{Kordyuk02,Kordyuk06a,Kordyuk10}. In this case, some natural questions are: (a)
whether the antibonding component of the electron quasiparticle excitation spectrum is completely independent
of the bonding component or not? (b) whether the behavior of the electron quasiparticle excitations determined
by the low-energy electronic structure is universal or not? (c) why the optimized $T_{\rm c}$ in the presence
of the bilayer coupling is enhanced?

The effect of the renormalization of the electrons in cuprate superconductors detected from ARPES experiments
is quantitatively characterized by the experimentally measurable quantities \cite{Damascelli03,Campuzano04,Fink07,Kordyuk14,Zhou18,Carbotte11,Kordyuk06a,Kordyuk10,Garcia10,Chatterjee06,He14}
such as the EFS reconstruction, the complicated line-shape in the electron quasiparticle excitation spectrum,
the kinks in the electron quasiparticle dispersion, and the ARPES autocorrelation. Recently, we
\cite{Feng15a,Feng16,Zhao17,Gao18,Gao18a,Gao19} have studied the renormalization of the electrons in the
single-layer cuprate superconductors based on the framework of the kinetic-energy driven SC mechanism, and
reproduced the main features observed from the corresponding ARPES experiments, including the renormalization
from the quasiparticle scattering reduces the spectral weight in EFS to the tips of the Fermi arcs
\cite{Zhao17,Gao18,Gao18a}, the charge-order correlation driven by the EFS instability with the characteristic
charge-order wave vector corresponding to the straight hot spots on EFS \cite{Feng16,Zhao17,Gao18}, the striking
PDH structure in the electron quasiparticle excitation spectrum \cite{Zhao17,Gao18,Gao18a}, and the remarkable
ARPES autocorrelation and its connection with the quasiparticle scattering interference (QSI) \cite{Gao19}.
However, the significant coupling effect between the copper-oxide layers on the renormalization of the electrons
has not been clarified in these discussions due to the limitation in the single-layer case. In this paper, we
study the characteristic features of the energy and momentum dependence of the renormalization of the electrons
in the bilayer cuprate superconductors along with this line by taking into account the bilayer coupling effect.
Our results show that the electron quasiparticle excitation spectrum is split into its bonding and antibonding
components due to the presence of the bilayer coupling, with each component that is independent, namely, two
copper-oxide layers within a unit cell are hybridized to form the bonding and antibonding layers, and then the
bonding layer is independent of the antibonding one, which leads to the bonding and antibonding EFS contours
deriving directly from the bonding and antibonding layers, respectively. However, in the underdoped and optimally
doped regimes, although the bonding (antibonding) EFS contour is truncated to form the disconnected bonding
(antibonding) Fermi arcs, the renormalization from the quasiparticle scattering further reduces almost all
spectral weight in the bonding (antibonding) Fermi arcs to the tips of the bonding (antibonding) Fermi arcs,
which in this case coincide with the bonding (antibonding) hot spots. These hot spots connected by the scattering
wave vectors ${\bf q}_{i}$ construct an {\it octet} scattering model, and then the enhancement of the quasiparticle
scattering processes with these scattering wave vectors ${\bf q}_{i}$ and the SC correlation are confirmed
via the result of the ARPES autocorrelation. This also indicates indirectly why the optimized $T_{\rm c}$ is
enhanced in the presence of the coupling between the copper-oxide layers within a unit cell. In a striking analogy
to the single-layer cuprate superconductors \cite{Zhao17,Gao18,Gao18a}, the remarkable PDH structure in the
antibonding (bonding) component of the electron quasiparticle excitation spectrum at around the antinodal and
nodal regions is caused by the corresponding peak structure in the antibonding (bonding) quasiparticle scattering
rate. However, this PDH structure at around the hot spots is developed mainly due to the pure bilayer coupling,
which is in a clear contrast to the single-layer case \cite{Zhao17,Gao18,Gao18a}. More specifically, our results
also show that although the kink in the electron quasiparticle dispersion is present all around EFS, when the
momentum moves away from the nodal region to the antinodal region, the dispersion kink becomes more pronounced,
and in particular, near the cut close to the antinode, develops into a break separating of the fasting dispersing
high-energy part of the electron quasiparticle excitation spectrum from the slower dispersing low-energy part.
Furthermore, the largest value of the kink energy appears along the nodal direction, and then decreases upon
approaching to the antinodal direction.

This paper is organized as follows. The basic formalism is presented in Sec. \ref{Formalism}, where we generalize
the electron normal and anomalous Green's functions obtained based on the kinetic-energy driven superconductivity
from the previous case in the single-layer cuprate superconductors \cite{Feng15a} to the present case in the bilayer
cuprate superconductors, and then evaluate explicitly the bondng and antibonding components of the electron
quasiparticle excitation spectrum. Within this basic formalism, we discuss the characteristic features of the
renormalization of the electrons in the bilayer cuprate superconductors in Sec. \ref{Renormalization}, where we
show clearly that the multiple electronic instabilities in the bilayer cuprate superconductors are driven by the
EFS instability, including charge order with the scattering that is peaked at the straight hot spots on EFS, and
then these multiple electronic instabilities coexist and compete with superconductivity. Finally, the summary and
discussions are presented in Sec. \ref{conclusions}. One Appendix is also included.

\section{Formalism}\label{Formalism}

The electron quasiparticle excitation spectrum $I({\bf k},\omega)\propto A({\bf k},\omega)$ can be probed by
the highly sophisticated ARPES experiments \cite{Damascelli03,Campuzano04,Fink07,Kordyuk14,Zhou18,Carbotte11},
where ${\bf k}$ is the in-plane momentum, $\omega$ is the energy of the initial state relative to the Fermi energy,
while the crucial electron spectral function $A({\bf k},\omega)$ is calculated in terms of the electron Green's
function starting from the microscopic model of the system. However, for the bilayer cuprate superconductors, the
electron quasiparticle excitation spectrum has been separated into its bonding and antibonding components due to the
presence of the bilayer coupling within a unit cell as  \cite{Feng-dl01,Chuang01,Eschrig06},
\begin{eqnarray}\label{QPE-spectrum}
I_{\nu}({\bf k},\omega)&=&|M_{\nu}({\bf k},\omega)|^{2}n_{\rm F}(\omega)A_{\nu}({\bf k},\omega),
\end{eqnarray}
where $\nu=1,2$ with $\nu=1$ ($\nu=2$) that represents the corresponding bonding (antibonding) component, the
dipole matrix element $M_{\nu}({\bf k}, \omega)$ is associated with the transition from the initial to final
electronic state, which can be affected by such things as incident photon energy and polarization as well as the
Brillouin zone (BZ) of the photoemitted electrons. However, it has been shown experimentally that this dipole
matrix element $M_{\nu}({\bf k},\omega)$ does not vary significantly with momentum and energy over the range of
the interest \cite{Damascelli03,Campuzano04,Fink07,Kordyuk14,Zhou18,Carbotte11}. In this case, as a qualitative
discussion in this paper, the magnitude of this dipole matrix element $M_{\nu}({\bf k},\omega)$ can be rescaled
to the unit. $n_{\rm F}(\omega)$ is the fermion distribution function, which indicates only occupied states are
probed in the ARPES experiments, while the electron spectral function $A_{\nu}({\bf k} ,\omega)$ in the SC-state
is related directly with the imaginary part of the electron normal Green's function $G_{\nu}({\bf k},\omega)$
as $A_{\nu}({\bf k},\omega)=-2{\rm Im}G_{\nu}({\bf k},\omega)$.

As we have mentioned above in Sec. \ref{Introduction}, the essential physics in cuprate superconductors occurs
in the copper-oxide layers \cite{Anderson87}. It is widely believed that a realistic model to provide an
acceptable description of the single copper-oxide layer is the $t$-$J$ model on a square lattice \cite{Anderson87,Zhang88,Anderson04,Lee06,Phillips10}. However, for the discussions of the bilayer coupling
effect on the renormalization of the electrons in the bilayer cuprate superconductors, the single-layer $t$-$J$
model should be extended by including the bilayer interaction as \cite{Lan13},
\begin{eqnarray}\label{bilayer-tJ-model}
H&=&-t\sum_{l\hat{\eta}a\sigma}C^{\dagger}_{la\sigma}C_{l+\hat{\eta}a\sigma}+t'\sum_{l\hat{\tau}a\sigma}
C^{\dagger}_{la\sigma}C_{l+\hat{\tau}a\sigma} \nonumber\\
&-&\sum_{la\neq b\sigma}t_{\perp}(l)C^{\dagger}_{la\sigma}C_{lb\sigma}+\mu\sum_{la\sigma}
C^{\dagger}_{la\sigma}C_{la\sigma} \nonumber\\
&+&J\sum_{l\hat{\eta}a}{\bf S}_{la}\cdot {\bf S}_{l+\hat{\eta}a}+J_{\perp}
\sum_{l}{\bf S}_{l1}\cdot {\bf S}_{l2},~~~~
\end{eqnarray}
where $a (b)=1,2$, is the copper-oxide layer index, the summation within the copper-oxide layer is over all
sites $l$, and for each $l$, over its nearest-neighbor (NN) sites $\hat{\eta}$ or the next NN sites
$\hat{\tau}$, $C^{\dagger}_{la\sigma}/C_{la\sigma}$ is electron operator that create/annihilate an electron
with spin $\sigma$ at lattice site $l$ in copper-oxide layer $a$,
${\bf S}_{la}=(S^{x}_{la},S^{y}_{la},S^{z}_{la})$ is spin operator, while $\mu$ is the chemical potential.
In particular, it has been found \cite{Feng-dl01,Chuang01,Eschrig06} that the observed bilayer splitting
can be described approximately by the momentum dependence of $t_{\perp}({\bf k})$ as,
\begin{eqnarray}\label{interlayer-hopping}
t_{\perp}({\bf k})={t_{\perp}\over 4}(\cos k_{x} -\cos k_{y})^{2},
\end{eqnarray}
which is expected for a coherent hopping between two copper-oxide layers within a unit cell. This
momentum-dependent form of the interlayer hopping (\ref{interlayer-hopping}) is strongly anisotropic
and also follows from the theoretical prediction
\cite{Massida88,Chakarvarty93,Andersen94,Andersen95,Liechtenstein96,Xiang96}. Moreover, the magnetic
exchange $J_{\perp}$ between two copper-oxide layers can be obtained in terms of the magnetic exchange $J$
within the copper-oxide layer as \cite{Lan13} $J_{\perp}=(t_{\perp}/t)^{2}J$. The no double electron
occupancy local constraint in this bilayer $t$-$J$ model (\ref{bilayer-tJ-model}) is ensured by
$\sum_{\sigma}C_{la\sigma}^{\dagger}C_{la\sigma}\leq 1$, which can be treated properly in analytical
calculations within the fermion-spin theory \cite{Feng9404,Feng15}, where the constrained electron
operators are decoupled as $C_{la\uparrow}=h^{\dagger}_{la\uparrow}S^{-}_{la}$ and
$C_{la\downarrow}=h^{\dagger}_{la\downarrow}S^{+}_{la}$, with the spinful fermion operator
$h_{la\sigma}= e^{-i\Phi_{la\sigma}}h_{la}$ that carries the charge of the electron together with some
effects of the spin configuration rearrangements due to the presence of the doped charge carrier itself,
while the spin operator $S_{la}$ carries the spin of the electron, then the no double electron occupancy
local constraint is satisfied in analytical calculations. In this formalism,
$\delta=\langle h^{\dagger}_{la\sigma} h_{la\sigma}\rangle =\langle h^{\dagger}_{la} h_{la}\rangle$
gives the total charge-carrier doping concentration.

In this fermion-spin representation, the bilayer $t$-$J$ model (\ref{bilayer-tJ-model}) can be rewritten
as,
\begin{eqnarray}\label{CSS-bilayer-tJ-model}
H&=&t\sum_{l\hat{\eta}a}(h^{\dagger}_{l+\hat{\eta}a\uparrow}h_{la\uparrow}S^{+}_{la}S^{-}_{l+\hat{\eta}a}
+h^{\dagger}_{l+\hat{\eta}a\downarrow} h_{la\downarrow}S^{-}_{la}S^{+}_{l+\hat{\eta}a})\nonumber\\
&-t'&\sum_{l\hat{\tau}a}(h^{\dagger}_{l+\hat{\tau}a\uparrow}h_{la\uparrow}S^{+}_{la}S^{-}_{l+\hat{\tau}a}
+h^{\dagger}_{l+\hat{\tau}a\downarrow} h_{la\downarrow}S^{-}_{la}S^{+}_{l+\hat{\tau}a})\nonumber \\
&+&\sum_{la\neq b}t_{\perp}(l)(h^{\dagger}_{lb\uparrow}h_{la\uparrow}S^{+}_{la}S^{-}_{lb}
+h^{\dagger}_{lb\downarrow}h_{la\downarrow} S^{-}_{la} S^{+}_{lb})\nonumber\\
&-&\mu\sum_{la\sigma}h^{\dagger}_{la\sigma}h_{la\sigma}+{J_{\rm eff}}\sum_{l\hat{\eta}a}{\bf S}_{la}
\cdot {\bf S}_{l+\hat{\eta}a} \nonumber\\
&+&{J_{\rm eff\perp}} \sum_{l}{\bf S}_{l1}\cdot {\bf S}_{l2},~~~~
\end{eqnarray}
with $J_{\rm eff}=J(1-\delta)^{2}$, $J_{\rm eff\perp}=J_{\perp}(1-\delta)^{2}$. In the following discussions,
the parameters in this bilayer $t$-$J$ model are chosen as $t/J=2.5$, $t'/t=0.28$, and $t_{\perp}/t=0.3$.
However, when necessary to compare with the experimental data, we take $J=100$ meV, which is the typical
value of the bilayer cuprate superconductors
\cite{Damascelli03,Campuzano04,Fink07,Kordyuk14,Zhou18,Carbotte11}.

For a microscopic description of the SC-state of cuprate superconductors, the kinetic-energy-driven SC
mechanism has been established based on the $t$-$J$ model in the fermion-spin representation
\cite{Feng15,Feng0306,Feng12}, where the interaction between charge carriers directly from the kinetic
energy by the exchange of spin excitations generates the formation of the d-wave charge-carrier pairs,
while the d-wave electron pairs originate from the d-wave charge-carrier pairing state are due to the
charge-spin recombination, and then these electron pairs condense into the d-wave SC-state. However, for
the bilayer cuprate superconductors, there are two coupled copper-oxide layers within a unit cell. In this
case, the charge-carrier normal and anomalous Green's functions are matrices \cite{Lan13}, and can be
expressed as, $\tilde{g}({\bf k},\omega)=g_{\rm L}({\bf k}, \omega)+g_{\rm T} ({\bf k},\omega)\sigma_{x}$
and $\tilde{\Gamma}^{\dagger}({\bf k},\omega)=\Gamma^{\dagger}_{\rm L}({\bf k},\omega)+
\Gamma^{\dagger}_{\rm T}({\bf k}, \omega) \sigma_{x}$, respectively, where $\sigma_{x}$ is the Pauli
matrix, $g_{\rm L}({\bf k},\omega)$ [$\Gamma^{\dagger}_{\rm L} ({\bf k},\omega)$] and
$g_{\rm T}({\bf k},\omega)$ [$\Gamma^{\dagger}_{\rm T}({\bf k},\omega)$] are the corresponding longitudinal
and transverse parts of the charge-carrier normal (anomalous) Green's function, respectively. Within the
framework of the kinetic-energy-driven SC mechanism, these longitudinal and transverse components of the
full charge-carrier normal and anomalous Green's functions for the bilayer $t$-$J$ model
(\ref{CSS-bilayer-tJ-model}) have been evaluated, and are given explicitly in Ref. \onlinecite{Lan13}.

In order to discuss the electronic properties, we need to calculate the electron normal and anomalous
Green's functions $\tilde{G}({\bf k},\omega)=G_{\rm L}({\bf k},\omega)+G_{\rm T}({\bf k},\omega)\sigma_{x}$
and $\tilde{\Im}^{\dagger}({\bf k},\omega)= \Im^{\dagger}_{\rm L} ({\bf k}, \omega)
+\Im^{\dagger}_{\rm T}({\bf k},\omega)\sigma_{x}$, which are the convolutions of the spin Green's function
and the corresponding charge-carrier normal and anomalous Green's functions in the fermion-spin representation.
However, in corresponding to the bonding and antibonding components of the electron quasiparticle excitation
spectrum in Eq. (\ref{QPE-spectrum}), these electron normal and anomalous Green's functions can be also
rewritten in the bonding-antibonding representation as,
$G_{\nu}({\bf k},\omega)=G_{\rm L}({\bf k},\omega)+(-1)^{\nu+1}G_{\rm T} ({\bf k}, \omega)$ and
$\Im^{\dagger}_{\nu}({\bf k},\omega)=\Im^{\dagger}_{\rm L}({\bf k},\omega)
+(-1)^{\nu+1}\Im^{\dagger}_{\rm T} ({\bf k}, \omega)$. Following the kinetic-energy-driven SC mechanism, a
full charge-spin recombination scheme has been developed recently \cite{Feng15a}, where a charge carrier and
a localized spin are fully recombined into a physical electron. According to this full charge-spin
recombination scheme, we \cite{Feng15a} have obtained the electron normal and anomalous Green's functions of
the single layer $t$-$J$ model in the fermion-spin representation. In particular, these electron normal and
anomalous Green's functions of the single-layer $t$-$J$ model have been employed to discuss the renormalization
of the electrons in the single-layer cuprate superconductors \cite{Feng15a,Feng16,Zhao17,Gao18,Gao18a,Gao19}, and
the obtained results are well consistent with the corresponding experimental results. Following these previous
discussions for the single-layer case \cite{Feng15a,Feng16,Zhao17,Gao18,Gao18a,Gao19}, the electron normal and
anomalous Green's functions of the bilayer $t$-$J$ model (\ref{CSS-bilayer-tJ-model}) in the bonding-antibonding
representation can be evaluated explicitly as [see Appendix \ref{bonding-antibonding-Green-functions}],
\begin{widetext}
\begin{subequations}\label{EGFS}
\begin{eqnarray}
G_{\nu}({\bf k},\omega)&=&{1\over\omega-\varepsilon^{(\nu)}_{\bf k}-\Sigma_{\rm ph}^{(\nu)}({\bf k},\omega)
-[\Sigma_{\rm pp}^{(\nu)}({\bf k}, \omega)]^{2}/[\omega+\varepsilon^{(\nu)}_{\bf k}
+\Sigma_{\rm ph}^{(\nu)}({\bf k},-\omega)]},~~~~~\label{NEGF}\\
\Im^{\dagger}_{\nu}({\bf k},\omega)&=&{-\Sigma_{\rm pp}^{(\nu)}({\bf k},\omega)\over
[\omega-\varepsilon^{(\nu)}_{\bf k}-\Sigma_{\rm ph}^{(\nu)} ({\bf k},\omega)][\omega+\varepsilon^{(\nu)}_{\bf k}
+\Sigma_{\rm ph}^{(\nu)}({\bf k},-\omega)]-[\Sigma_{\rm pp}^{(\nu)}({\bf k},\omega)]^{2}}, \label{ANEGF}
\end{eqnarray}
\end{subequations}
\end{widetext}
where $\varepsilon^{(\nu)}_{\bf k}=\varepsilon_{\bf k}+(-1)^{\nu}t_{\perp}({\bf k})$ is the bare dispersion
relation, with $\varepsilon_{\bf k}=-Zt \gamma_{\bf k}+Zt'\gamma'_{\bf k}+\mu$,
$\gamma_{\bf k}=(\cos{{\rm k}_x}+\cos{{\rm k}_y})/2$, $\gamma'_{\bf k}=\cos{{\rm k}_x}\cos{{\rm k}_y}$, and
the number of the NN or next NN sites $Z$, while $\Sigma_{\rm ph}^{(\nu)}({\bf k},\omega)$ and
$\Sigma_{\rm pp}^{(\nu)}({\bf k},\omega)$ are the electron self-energies in the particle-hole and
particle-particle channels, respectively, which are the generally complex quantities describing the influence
of the interaction between the electrons mediated by the exchange of spin excitations on the propagation of
the electron quasiparticle, and are evaluated explicitly in terms of the full charge-spin reconbination in
Appendix \ref{bonding-antibonding-Green-functions}.

According to the above electron normal Green's function in Eq. (\ref{NEGF}), the electron spectral function
$A_{\nu}({\bf k},\omega)=-2{\rm Im} G_{\nu}({\bf k},\omega)$ in the bonding-antibonding representation can
be evaluated explicitly as,
\begin{eqnarray}\label{ESF}
A_{\nu}({\bf k},\omega)={2\Gamma_{\nu}({\bf k},\omega)\over [\omega-\bar{E}_{\nu}({\bf k},\omega)]^{2}
+\Gamma^{2}_{\nu}({\bf k},\omega)},
\end{eqnarray}
with the corresponding quasiparticle scattering rate $\Gamma_{\nu}({\bf k},\omega)$ and the renormalized
band structures (then the renormalized dispersion relation) $\bar{E}_{\nu}({\bf k},\omega)$ that can be
expressed explicitly as,
\begin{widetext}
\begin{subequations}\label{MESE}
\begin{eqnarray}
\Gamma_{\nu}({\bf k},\omega)&=&\left |{\rm Im}\Sigma^{(\nu)}_{\rm ph}({\bf k},\omega)
-{[\Sigma^{(\nu)}_{\rm pp}({\bf k},\omega)]^{2}{\rm Im} \Sigma^{(\nu)}_{\rm ph}({\bf k},-\omega)\over
[\omega+\varepsilon^{(\nu)}_{\bf k}+{\rm Re}\Sigma^{(\nu)}_{\rm ph}({\bf k},-\omega)]^{2}+[{\rm Im}
\Sigma^{(\nu)}_{\rm ph}({\bf k},-\omega)]^{2}}\right |, ~~~~~\label{EQDSR}\\
\bar{E}_{\nu}({\bf k},\omega)&=&\varepsilon^{(\nu)}_{\bf k}
+{\rm Re}\bar{\Sigma}^{(\nu)}_{\rm ph}({\bf k},\omega), ~~~\label{MRESE}\\
{\rm Re}\bar{\Sigma}^{(\nu)}_{\rm ph}({\bf k},\omega)&=&{\rm Re}\Sigma^{(\nu)}_{\rm ph}({\bf k},\omega)
+{[\Sigma^{(\nu)}_{\rm pp}({\bf k}, \omega)]^{2}[\omega+\varepsilon^{(\nu)}_{\bf k}
+{\rm Re}\Sigma^{(\nu)}_{\rm ph}({\bf k},-\omega)]\over [\omega+\varepsilon^{(\nu)}_{\bf k}
+{\rm Re} \Sigma^{(\nu)}_{\rm ph}({\bf k},-\omega)]^{2}
+[{\rm Im}\Sigma^{(\nu)}_{\rm ph}({\bf k}, -\omega)]^{2}}, ~~~\label{RESE}
\end{eqnarray}
\end{subequations}
\end{widetext}
respectively, where ${\rm Re}\Sigma^{(\nu)}_{\rm ph}({\bf k},\omega)$ and
${\rm Im}\Sigma^{(\nu)}_{\rm ph}({\bf k},\omega)$ are the real and imaginary parts of the electron
self-energy $\Sigma^{(\nu)}_{\rm ph}({\bf k},\omega)$, respectively. Substituting this electron spectral
functions $A_{\nu}({\bf k},\omega)$ into Eq. (\ref{QPE-spectrum}), we thus obtain the quasiparticle
excitation spectrum $I_{\nu}({\bf k},\omega)$. In this case, the energy and lifetime renormalization of
the electrons in the bilayer cuprate superconductors are directly described by the renormalized dispersion
relation $\bar{E}_{\nu}({\bf k},\omega)$ and the quasiparticle scattering rate $\Gamma_{\nu}({\bf k},\omega)$,
respectively. This is why the ARPES spectral line shape can give direct access to the lifetime of the electron
quasiparticle excitation and can provide the insight into the nature of the underlying interaction between
the electrons \cite{Damascelli03,Campuzano04,Fink07,Kordyuk14,Zhou18,Carbotte11}.

\section{Quantitative characteristics of renormalization of electrons}\label{Renormalization}

In this section, we discuss the energy and momentum dependence of the renormalization of the electrons in the
presence of the bilayer coupling, and show how the bilayer interaction gives some additional effects on the
redistribution of the spectral weight on EFS, the PDH structure in the electron quasiparticle excitation
spectrum, the kink in the electron quasiparticle dispersion, and the ARPES autocorrelation in the bilayer
cuprate superconductors.

\subsection{Electron Fermi surface instability}\label{Fermi-arcs}

\begin{figure}[h!]
\centering
\includegraphics[scale=1.4]{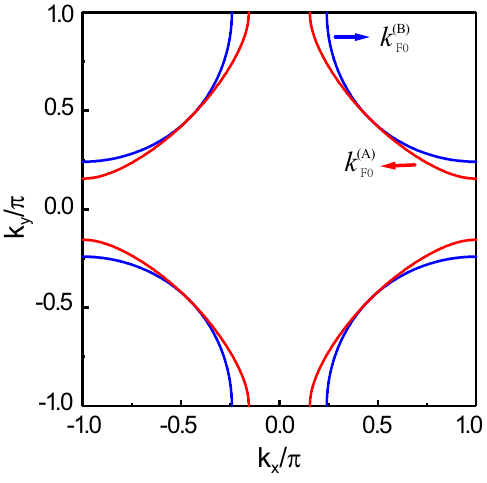}
\caption{(Color online) The bonding (blue line) and antibonding (red line) electron Fermi surface contours
obtained in the self-consistent renormalized mean-field level at $\delta=0.12$ with $T=0.002J$ for $t/J=2.5$,
$t'/t=0.28$, and $t_{\perp}/t=0.3$.  \label{MF-spectral-maps}}
\end{figure}

EFS is the zero-energy contour in momentum-space that separates the filled electronic states from the empty
electronic states. Superconductivity is one of several phenomena, including different ordered electronic
states \cite{Timusk99,Hufner08,Comin16,Vishik18}, that arise from the interaction of electrons near EFS
\cite{Carbotte11,Choi18,Bok16,Maier08,Scalapino12,Mou19}. The EFS topology is therefore crucial to the
understanding of these phenomena and their relationships. Experimentally, the intensity of ARPES spectra at
zero energy is usually used to map out the underlying EFS \cite{Damascelli03,Campuzano04,Fink07,Kordyuk14,Zhou18},
i.e., the underlying EFS is determined by looking at the electron quasiparticle excitation spectrum
$I({\bf k},\omega=0)$ to map out the locus of the maximum in the intensity of $I({\bf k},\omega=0)$. However,
as shown in Eq. (\ref{QPE-spectrum}), the electron quasiparticle excitation spectrum $I({\bf k},\omega)$ in the
bilayer cuprate superconductors has been separated into the corresponding bonding and antibonding components due
to the presence of the bilayer coupling \cite{Feng-dl01,Chuang01,Eschrig06}, which complicates the physical
properties of EFS. For instance, in the self-consistent renormalized mean-field (MF) level
[see, Eq. (\ref{Normal-MF-ESF})], the bonding and antibonding components of the electron quasiparticle excitation
spectrum in the {\it normal-state} consist of the delta function peaks located at the precise energy and
momentum given by the self-consistent renormalized MF band structure
$I^{(0)}_{1}({\bf k},\omega)=2\pi n_{\rm F}(\omega)Z^{(1)}_{\rm F}\delta(\omega-\bar{\varepsilon}^{(1)}_{\bf k})$
and $I^{(0)}_{2}({\bf k},\omega)=2\pi n_{\rm F}(\omega)Z^{(2)}_{\rm F}\delta(\omega-\bar{\varepsilon}^{(2)}_{\bf k})$,
respectively, and then the intensities of $I^{(0)}_{1}({\bf k},0)$ and $I^{(0)}_{2}({\bf k},0)$ at zero
binding-energy can be used to map out the underlying EFS, which is plotted in Fig. \ref{MF-spectral-maps}, where
the doping concentration $\delta=0.12$ and temperature $T=0.002J$. In comparison with the corresponding
self-consistent renormalized MF result of the single closed EFS contour in the single-layer case
\cite{Zhao17,Gao18}, there are two closed EFS contours ${\bf k}^{\rm (B)}_{\rm F0}$ and
${\bf k}^{\rm (A)}_{\rm F0}$ in the bilayer case deriving directly from the bonding and antibonding layers.
These bonding and antibonding EFS contours correspond to different charge-carrier doping levels \cite{Liu92,Kordyuk04,Chuang04,Borisenko04,Loret18,Ai19,Gao-Q19}. The result in Fig. \ref{MF-spectral-maps}
therefore confirms that the existence of two closed bonding and antibonding EFS contours
${\bf k}^{\rm (B)}_{\rm F0}$ and ${\bf k}^{\rm (A)}_{\rm F0}$ are due to the presence of the bilayer coupling.
In particular, the peaks with the same height distribute uniformly along with both the bonding and antibonding
EFS contours, reflecting a fact that these two closed bonding and antibonding EFS contours
${\bf k}^{\rm (B)}_{\rm F0}$ and ${\bf k}^{\rm (A)}_{\rm F0}$ in the bilayer case are necessarily two surfaces
in momentum-space on which the electron lifetime becomes infinitely long in the limit as one approaches these
bonding and antibonding EFS contours. However, according to one of the self-consistent equations
[see, Eq. (\ref{SCE-3})], the effective area of these EFS contours contains $1-\delta$ electrons, and therefore
it fulfills Luttinger's theorem \cite{Luttinger60}. Moreover, in corresponding to the form of the interlayer
coherent hopping in Eq. (\ref{interlayer-hopping}), the separation between the bonding and antibonding EFS
contours ${\bf k}^{\rm (B)}_{\rm F0}$ and ${\bf k}^{\rm (A)}_{\rm F0}$ has a largest value at around the
antinodes, and then it smoothly decreases with the move of the momentum away from the antinodal region,
eventually disappearing at around the nodes. In the bilayer cuprate superconductors, two copper-oxide layers
in the unit cell are equivalent, and they are hybridized. However, these self-consistent renormalized MF
results also show that the electrons in the bonding and antibonding bands have different EFS contours except
for at around the nodes, in other words, the bonding layer is independent completely of the antibonding layer.

\begin{figure}[h!]
\centering
\includegraphics[scale=1.4]{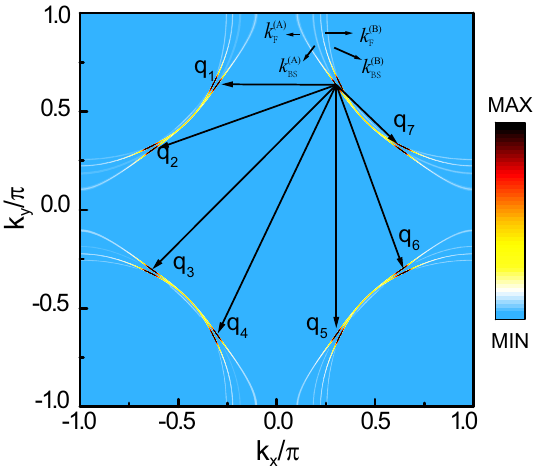}
\caption{(Color online) The intensity map of the bonding and antibonding components of the electron
quasiparticle excitation spectrum $I_{1}({\bf k},0)$ and $I_{2}({\bf k},0)$ at $\delta=0.12$ with
$T=0.002J$ for $t/J=2.5$, $t'/t=0.28$, and $t_{\perp}/t=0.3$. \label{spectral-maps}}
\end{figure}

However, the spectral weights in the bonding and antibonding components of the electron quasiparticle
excitation spectrum $I_{1}({\bf k},\omega)$ and $I_{2}({\bf k},\omega)$ in Eq. (\ref{QPE-spectrum}) are
redistributed strongly by the inclusion of the energy and momentum dependence of the bonding and antibonding
electron self-energies $\Sigma_{\rm ph}^{(1)}({\bf k},\omega)$ and $\Sigma_{\rm ph}^{(2)}({\bf k},\omega)$,
respectively, which captures the strong interaction between the electrons by the exchange of spin excitations
in the bilayer cuprate superconductors. To see this strong redistribution of the spectral weights in the
bonding and antibonding EFS contours more clearly, we plot the map from both
$I_{1}({\bf k},0)$ and $I_{2}({\bf k},0)$ in Eq. (\ref{QPE-spectrum}) at zero binding-energy for
$\delta=0.12$ with $T=0.002J$ in Fig. \ref{spectral-maps}. Apparently, the quasiparticle peaks gain a finite
width in a way dependent on the momentum of the electron quasiparticle excitations, where the most
characteristic features can be summarized as: (a) with respect to the self-consistent renormalized MF result
in Fig. \ref{MF-spectral-maps}, a bifurcation for each EFS contour occurs at the Fermi energy level, i.e.,
a bonding (antibonding) sheet labelled as ${\bf k}^{\rm (B)}_{\rm BS}$ (${\bf k}^{\rm (A)}_{\rm BS}$) is
bifurcated from the bonding (antibonding) EFS contour ${\bf k}^{\rm (B)}_{\rm F}$ (${\bf k}^{\rm (A)}_{\rm F}$)
except for at around the special points ${\bf k}^{\rm (B)}_{\rm HS}$ (${\bf k}^{\rm (A)}_{\rm HS}$), where
this bifurcation disappears; (b) however, the spectral weights on these bonding and antibonding sheets
${\bf k}^{\rm (B)}_{\rm BS}$ and ${\bf k}^{\rm (A)}_{\rm BS}$ are suppressed, leading to that these two
sheets become unobservable in experiments \cite{Loret18,Ai19,Gao-Q19}; (c) on the other hand, the lifetime
renormalization of the electrons on the bonding (antibonding) EFS contour ${\bf k}^{\rm (B)}_{\rm F}$
(${\bf k}^{\rm (A)}_{\rm F}$) is angular dependent, i.e., the spectral weight in the bonding (antibonding)
component of the electron quasiparticle excitation spectrum on ${\bf k}^{\rm (B)}_{\rm F}$
(${\bf k}^{\rm (A)}_{\rm F}$) at around the bonding (antibonding) antinodal region is suppressed, and then
the bonding (antibonding) EFS contour is truncated to form the disconnected bonding (antibonding) Fermi arcs \cite{Norman98,Kanigel06,Tanaka06,Kanigel07,Meng11,Ideta12,Kondo13} located around the nodal region with the
tips of the bonding (antibonding) Fermi arcs that are just corresponding to these special points
${\bf k}^{\rm (B)}_{\rm HS}$ (${\bf k}^{\rm (A)}_{\rm HS}$); (d) more specifically, the points with the
strongest intensity on the bonding (antibonding) Fermi arcs do not reach the nodes, but instead reach exactly
these special points ${\bf k}^{\rm (B)}_{\rm HS}$ (${\bf k}^{\rm (A)}_{\rm HS}$), and only in this sense,
these special points ${\bf k}^{\rm (B)}_{\rm HS}$ (${\bf k}^{\rm (A)}_{\rm HS}$) are called as the hot spots
on the bonding (antibonding) EFS contour. To see this remarkable feature more clearly, we plot the intensity
map of $I_{1}({\bf k},0)$ and $I_{2}({\bf k},0)$ in the first quarter of BZ at zero binding-energy for
$\delta=0.12$ with $T=0.002J$ in Fig. \ref{spectral-maps-3D}, where although the corresponding spectral
densities on the bonding and antibonding Fermi arcs are almost the same, the highest peak heights at the
bonding (antibonding) Fermi arcs, marked by the black (red) circles, reach exactly the bonding (antibonding)
hot spots. In other words, the renormalization of the quasiparticle scattering reduces almost all spectral
weights in the bonding (antibonding) Fermi arcs to the bonding (antibonding) hot spots, which conform well
to the ARPES experiments \cite{Loret18,Chatterjee06,He14,Sassa11,Kaminski05}, where the observed result in
the bilayer cuprate superconductors indicates that the sharp quasiparticle peaks with largest spectral weight
appear always at off-node place; (e) in the underdoped and optimally doped regimes, the strong redistribution
of the spectral weight in the antibonding EFS contour is completely independent of the strong redistribution
of the spectral weight in the bonding EFS contour; (f) however, the bilayer splitting between the bonding and
antibonding bands in Eq. (\ref{interlayer-hopping}) is rather weak around the nodal region, and disappears
exactly at around the nodes. This leads to that the bonding and antibonding Fermi arcs in each quarter of BZ
are almost degenerate around the nodal region except for at around the hot spots. In this case, the boundary
between the bonding and antibonding Fermi arcs around the nodal region is too dim to be observed. This is also
why only single Fermi arc can be observed experimentally in each quarter of BZ in the underdoped and optimally
doped bilayer cuprate superconductors \cite{Norman98,Kanigel06,Tanaka06,Kanigel07,Meng11,Ideta12,Kondo13}.
In particular, in the heavily overdoped regime, the renormalization of the electrons within each copper-oxide
layer is weakened, and then as in the self-consistent renormalized MF case, the bonding (antibonding)
Fermi arcs cover the full length of the bonding (antibonding) EFS contour, leading to that two closed EFS
contours induced mainly by the bilayer coupling are observed experimentally. Furthermore, we have also studied
the redistribution of the spectral weights in the bonding and antibonding EFS contours in the normal-state,
and the results indicate that this unusual redistribution of the spectral weights in the bonding and
antibonding EFS contours appeared in the SC-state also persists into the normal-state.

\subsection{Octet scattering model and related coexistence and competition between multiple electronic orders
and superconductivity}\label{charge-order}

\begin{figure}[h!]
\centering
\includegraphics[scale=0.5]{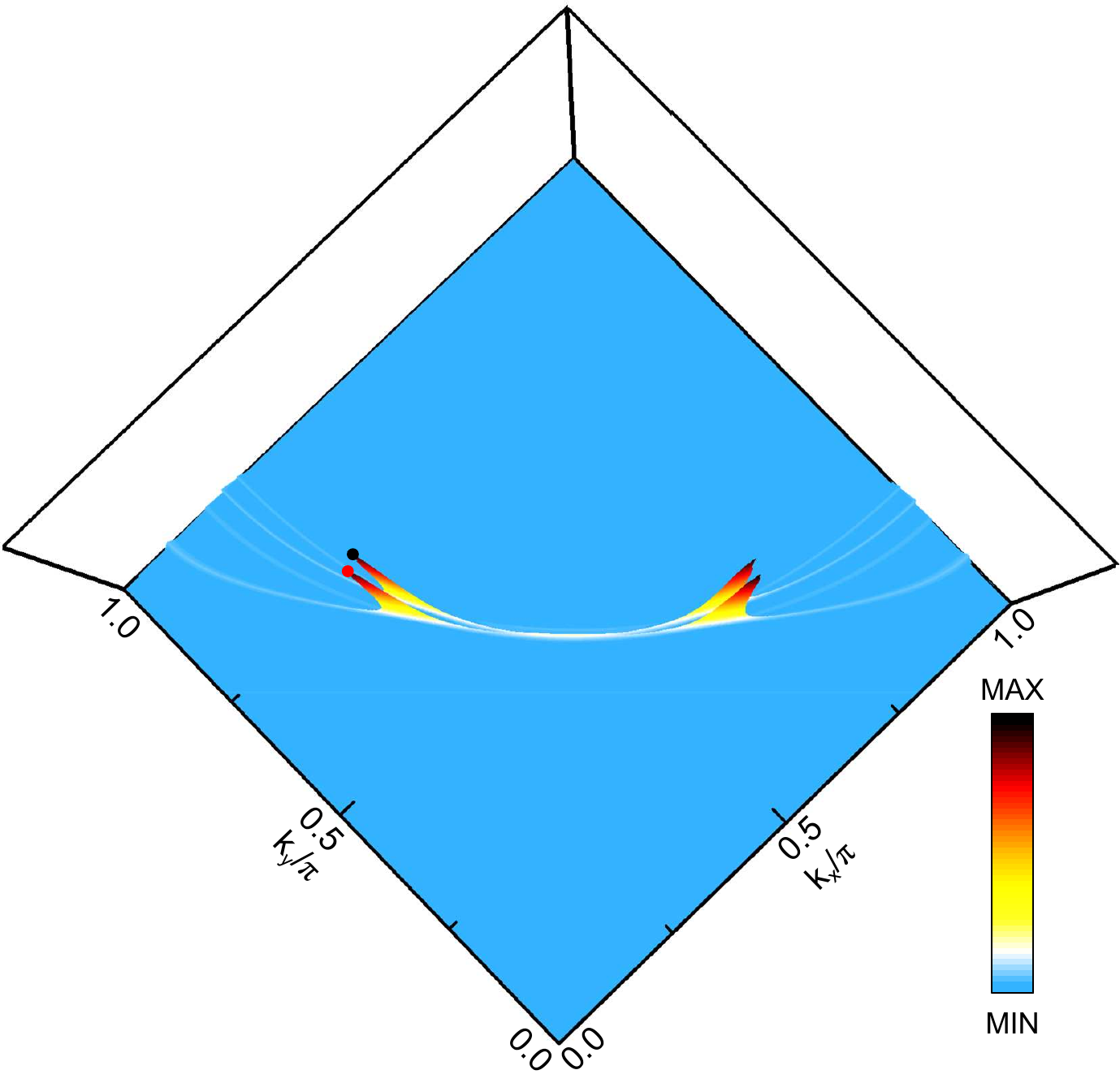}
\caption{(Color online) The spectral-weight distribution in the bonding and antibonding components of the
electron quasiparticle excitation spectrum $I_{1}({\bf k},0)$ and $I_{2}({\bf k},0)$ in the first quarter of
the Brillouin zone at $\delta=0.12$ with $T=0.002J$ for $t/J=2.5$, $t'/t=0.28$, and $t_{\perp}/t=0.3$. The
black circle indicates the location of the bonding hot spot, while the red circle indicates the location
of the antibonding hot spot. \label{spectral-maps-3D}}
\end{figure}

Since the renormalization from the quasiparticle scattering reduces almost all spectral weights in the
bonding and antibonding EFS contours ${\bf k}^{\rm (B)}_{\rm F}$ and ${\bf k}^{\rm (A)}_{\rm F}$ to the
bonding and antibonding hot spots ${\bf k}^{\rm (B)}_{\rm HS}$ and ${\bf k}^{\rm (A)}_{\rm HS}$, respectively,
these hot spots connected by the scattering wave vectors ${\bf q}_{i}$ shown in Fig. \ref{spectral-maps}
construct an {\it octet} scattering model \cite{Chatterjee06,He14,Gao19}, which also reflects a fact that the
hot spots are thus special points on the EFS contours that are displaced from each other by the scattering
wave vectors ${\bf q}_{i}$. In this case, a bewildering variety of competing phases described by the
quasiparticle scattering processes with the scattering wave vectors ${\bf q}_{i}$ therefore are driven by
the EFS instability. This is also why a remarkable phenomenon of cuprate superconductor is the coexistence
and competition between the multiple nearly-degenerate electronic orders or instabilities and superconductivity \cite{Timusk99,Hufner08,Comin16,Vishik18}. Furthermore, in comparison with the corresponding case in the
single-layer cuprate superconductors \cite{Gao18,Gao18a,Gao19}, the quasiparticle scattering processes with
the scattering wave vectors ${\bf q}_{i}$ and the SC correlation are enhanced in the bilayer cuprate
superconductors. This follows a fact that for a given quasiparticle scattering process, there are four
corresponding scattering wave vectors: the scattering wave vector ${\bf q}^{\rm (AA)}_{i}$ connecting between
the antibonding hot spots, the scattering wave vector ${\bf q}^{\rm (AB)}_{i}$ connecting between the
antibonding and bonding hot spots, the scattering wave vector ${\bf q}^{\rm (BA)}_{i}$ connecting between the
bonding and antibonding hot spots, and the scattering wave vector ${\bf q}^{\rm (BB)}_{i}$ connecting between
the bonding hot spots. This given quasiparticle scattering process with the corresponding four scattering
wave vectors ${\bf q}^{\rm (AA)}_{i}$, ${\bf q}^{\rm (AB)}_{i}$, ${\bf q}^{\rm (BA)}_{i}$, and
${\bf q}^{\rm (BB)}_{i}$ contributes to the same electronic instability, and then this electronic instability
is enhanced. In particular, the special quasiparticle scattering process with the scattering wave vector
${\bf q}_{1}$ connecting the hot spots on the straight Fermi arcs shown in Fig. \ref{spectral-maps} describes
the pairing of electrons and holes at ${\bf k}\uparrow$ and ${\bf k}+{\bf Q}^{(\alpha\alpha')}_{\rm HS}\uparrow$,
with ${\bf Q}^{\rm (AA)}_{\rm HS} ={\bf q}^{\rm (AA)}_{1}$, ${\bf Q}^{\rm (AB)}_{\rm HS}={\bf q }^{\rm (AB)}_{1}$,
${\bf Q}^{\rm (BA)}_{\rm HS}={\bf q}^{\rm (BA)}_{1}={\bf Q}^{\rm (AB)}_{\rm HS}$, and
${\bf Q}^{\rm (BB)}_{\rm HS }={\bf q}^{\rm (BB)}_{1}$, which is responsible for the charge-order formation
\cite{Timusk99,Hufner08,Comin16,Vishik18,Comin14,Wu11,Chang12,Ghiringhelli12,Neto14,Fujita14,Campi15,Comin15a,Peng16,Hinton16}.
On the other hand, the electron pairing at ${\bf k}+{\bf Q}^{(\alpha\alpha')}_{\rm HS}\uparrow$
and $-{\bf k}-{\bf Q} ^{(\alpha\alpha')}_{\rm HS}\downarrow$ states related with the wave vector ${\bf q}_{4}$
shown in Fig. \ref{spectral-maps} is responsible for the SC correlation \cite{Gao18,Hinton16}. These two special
processes therefore indicate a coexistence and competition between charge order and superconductivity just as
they have been observed from a wide variety of measurement techniques
\cite{Timusk99,Hufner08,Comin16,Vishik18,Comin14,Wu11,Chang12,Ghiringhelli12,Neto14,Fujita14,Campi15,Comin15a,Peng16,Hinton16}.
Moreover, the average value of these four quasiparticle scattering wave vectors
${\bf Q}_{\rm HS}=[{\bf Q}^{\rm (AA)}_{\rm HS}+ {\bf Q}^{\rm (AB)}_{\rm HS}+{\bf Q}^{\rm (BA)}_{\rm HS}
+{\bf Q}^{\rm (BB)}_{\rm HS}]/4$ at the doping concentration $\delta\sim 0.12$ is found to be
${\bf Q}_{\rm HS}\sim 0.31$ (here we use the reciprocal units), which is well consistent with experimental
average value \cite{Comin14} of the charge-order wave vector $Q_{\rm CO}\sim 0.256$ observed in the
underdoped bilayer cuprate superconductors Bi$_{2}$Sr$_{2-x}$La$_{x}$CuO$_{6+\delta}$. Superconductivity
in cuprate superconductors occurs on adding charge carriers to the copper-oxide layers. As a natural
consequence of this adding charge-carrier progression, the charge-order wave vector ${\bf Q}_{\rm HS}$ is
also doping dependent, where we find that as the doping concentration is increased, ${\bf Q}_{\rm HS}$
decreases almost linearly, also in qualitative agreement with the experimental results
\cite{Timusk99,Hufner08,Comin16,Vishik18,Comin14,Wu11,Chang12,Ghiringhelli12,Neto14,Fujita14,Campi15,Comin15a,Peng16,Hinton16}
of the doping dependence of the charge-order wave vector $Q_{\rm CO}$ observed from the bilayer cuprate
superconductor Bi$_{2}$Sr$_{2-x}$La$_{x}$CuO$_{6+\delta}$. The enhancement of the electronic orders and SC
correlation in the bilayer cuprate superconductors will be discussed further in
Sec. \ref{ARPES-autocorrelation}.

The essential ingredients to develop the redistribution of the spectral weights in the bonding and
antibonding EFS contours in the bilayer cuprate superconductors are the following two: the multiple
electronic orders as in the single-layer case \cite{Feng16,Zhao17,Gao18,Gao18a} and the bilayer coupling.
The electron self-energy in the particle-hole channel $\Sigma^{(\nu)}_{\rm ph}({\bf k},\omega)$ in
Eq. (\ref{EGFS}) in the bonding-antibonding representation can be also rewritten as \cite{Lan13},
\begin{eqnarray}\label{PG}
\Sigma^{(\nu)}_{\rm ph}({\bf k},\omega)\approx {[\bar{\Delta}^{(\nu)}_{\rm co}({\bf k})]^{2}
\over\omega+\varepsilon^{(\nu)}_{0{\bf k}}},
\end{eqnarray}
where the corresponding dispersion relation $\varepsilon^{(\nu)}_{0{\bf k}}$ and the momentum-dependent
gap $\bar{\Delta}^{(\nu)}_{\rm co}({\bf k})$ are obtained directly from the electron self-energy
$\Sigma^{(\nu)}_{\rm ph}({\bf k},\omega)$ in Eq. (\ref{EGFS}) and its antisymmetric part
$\Sigma^{(\nu)}_{\rm pho}({\bf k},\omega)$ as
$\varepsilon^{(\nu)}_{0{\bf k}}=-\Sigma^{(\nu)}_{\rm ph}({\bf k},0)/\Sigma^{(\nu)}_{\rm pho}({\bf k}, 0)$
and $\bar{\Delta}^{(\nu)}_{\rm co}({\bf k})
=\Sigma^{(\nu)}_{\rm ph}({\bf k},0)/\sqrt{-\Sigma^{(\nu)}_{\rm pho}({\bf k},0)}$, respectively.
Since this momentum-dependent gap $\bar{\Delta}^{(\nu)}_{\rm co}({\bf k})$ originates from the electron
self-energy $\Sigma^{(\nu)}_{\rm ph}({\bf k}, \omega)$ in the particle-hole channel, it can be identified
as being a region of the electron self-energy effect in the particle-hole channel in which (i) the
momentum-dependent gap suppresses the spectral weight of the electron quasiparticle excitation spectrum
around the antinodal region to form the Fermi arcs, as the role played by the pseudogap, and (ii) the
momentum-dependent gap reduces the spectral weight in the Fermi arcs to the hot spots to construct an octet
scattering model with the scattering wave vectors ${\bf q}_{i}$ connecting the hot spots, as the role played
by the electronic order gap \cite{LeBlanc14}, such as the charge-order gap \cite{Feng19}. This is why the
multiple nearly-degenerate electronic orders exist within the pseudogap phase, appearing below a temperature
$T_{\rm co}$ well above $T_{\rm c}$ in the underdoped and optimally doped regimes, and coexists with
superconductivity below $T_{\rm c}$. $T_{\rm co}$ is the temperature where the electronic order, such as
charge order, develops, and is of the order of the pseudogap crossover temperature $T^{*}$. This in turn
shows that the emergence of the electronic order gap $\bar{\Delta}^{(\nu)}_{\rm co}({\bf k} )$ is the
microscopic mechanism underlying multiple electronic orders driven by the EFS instability in cuprate
superconductors. The expression in Eq. (\ref{PG}) also shows that the imaginary part of
$\Sigma^{(\nu)}_{\rm ph}({\bf k},\omega)$ is related directly with the electronic order gap as,
\begin{eqnarray}\label{IESE}
{\rm Im}\Sigma^{(\nu)}_{\rm ph}({\bf k},\omega)\approx 2\pi[\bar{\Delta}^{(\nu)}_{\rm co}({\bf k})]^{2}
\delta(\omega+\varepsilon^{(\nu)}_{0{\bf k}}),
\end{eqnarray}
which therefore shows that the momentum dependence of the quasiparticle scattering rate
$\Gamma_{\nu}({\bf k},\omega)$ in Eq. (\ref{EQDSR}) is dominated by the momentum dependence of the
electronic order gap \cite{Hashimoto15} $\bar{\Delta}^{(\nu)}_{\rm co}({\bf k})$.

With the help of the above expression of the electron self-energy $\Sigma^{(\nu)}_{\rm ph}({\bf k},\omega)$
in Eq. (\ref{PG}), the electron normal and anomalous Green's functions in Eq. (\ref{EGFS}) in the SC-state
with the coexisting electronic orders can be reexpressed explicitly as,
\begin{subequations}\label{EGF1}
\begin{eqnarray}
G_{\nu}({\bf k},\omega)&=&\left ({U^{(\nu)2}_{1{\bf k}}\over\omega-E^{(\nu)}_{1{\bf k}}}
+{V^{(\nu)2}_{1{\bf k}}\over\omega+E^{(\nu)}_{1{\bf k}}} \right )\nonumber\\
&+&\left ({U^{(\nu)2}_{2{\bf k}}\over\omega-E^{(\nu)}_{2{\bf k}}} +{V^{(\nu)2}_{2{\bf k}}\over\omega
+E^{(\nu)}_{2{\bf k}}}\right ), \label{DEGF1}\\
\Im^{\dagger}_{\nu}({\bf k},\omega)&=&-{a^{(\nu)}_{1{\bf k}}\bar{\Delta}^{(\nu)}({\bf k})\over
2E^{(\nu)}_{1{\bf k}}}\left ({1\over\omega- E^{(\nu)}_{1{\bf k}}}-{1\over\omega + E^{(\nu)}_{1{\bf k}}}
\right )\nonumber\\
&+& {a^{(\nu)}_{2{\bf k}}\bar{\Delta}^{(\nu)}({\bf k})\over 2E^{(\nu)}_{2{\bf k}}}\left ({1\over\omega
-E^{(\nu)}_{2{\bf k}}}-{1\over\omega+ E^{(\nu)}_{2{\bf k}}} \right ), ~~~~~~~~~ \label{ODEGF1}
\end{eqnarray}
\end{subequations}
where $a^{(\nu)}_{1{\bf k}}=(E^{(\nu)2}_{1{\bf k}}-\varepsilon^{(\nu)2}_{0{\bf k}})/(E^{(\nu)2}_{1{\bf k}}
-E^{(\nu)2}_{2{\bf k}})$, $a^{(\nu)}_{2{\bf k}}=(E^{(\nu)2}_{2{\bf k}}-
\varepsilon^{(\nu)2}_{0{\bf k}})/(E^{(\nu)2}_{1{\bf k}}-E^{(\nu)2}_{2{\bf k}})$, the electron pair gap
$\bar{\Delta}^{(\nu)}({\bf k})=\Sigma_{\rm pp}^{(\nu)}({\bf k},\omega=0)=\bar{\Delta}_{\rm L}
\gamma^{\rm (d)}_{\bf k}+(-1)^{1+\nu}\bar{\Delta}_{\rm T}$, with
$\gamma^{\rm (d)} _{\bf k} =[\cos k_{x}-\cos k_{y} ]/2$. However, as we have mentioned above, the
electron quasiparticle excitation spectrum has been split into its bonding and antibonding components
by the bilayer coupling, with each component that is independent. In this case, the bonding electron
quasiparticle excitations in the SC-state with the coexisting electronic orders become superpositions
of four bonding-eigenstates with the corresponding four bonding-energy eigenvalues $E^{(1)}_{1{\bf k}}$,
$-E^{(1)}_{1{\bf k}}$, $E^{(1)}_{2{\bf k}}$, and $-E^{(1)}_{2{\bf k}}$, while the antibonding electron
quasiparticle excitations in the SC-state with the coexisting electronic orders become superpositions of
four antibonding-eigenstates with the corresponding four antibonding-energy eigenvalues $E^{(2)}_{1{\bf k}}$,
$-E^{(2)}_{1{\bf k}}$, $E^{(2)}_{2{\bf k}}$, and $-E^{(2)}_{2{\bf k}}$, where the energy eigenvalues
$E^{(\nu)}_{1{\bf k}}=\sqrt{[K^{(\nu)}_{1{\bf k} }+K^{(\nu)}_{2{\bf k}}]/2}$,
$E^{(\nu)}_{2{\bf k}}=\sqrt{[K^{(\nu)}_{1{\bf k} }-K^{(\nu)}_{2{\bf k}} ]/2}$, and the kernel functions,
\begin{eqnarray*}
K^{(\nu)}_{1{\bf k}}&=&\varepsilon^{(\nu)2}_{\bf k}+\varepsilon^{(\nu)2}_{0{\bf k}}
+2\bar{\Delta}^{(\nu)2}_{\rm co}({\bf k})+ \bar{\Delta}^{(\nu)2} ({\bf k}),\\
K^{(\nu)}_{2{\bf k}}&=&\sqrt{(\varepsilon^{(\nu)2}_{\bf k}-\varepsilon^{(\nu)2}_{0{\bf k}})b^{(\nu)}_{1{\bf k}}
+4\bar{\Delta}^{(\nu)2}_{\rm co}({\bf k} )b^{(\nu)}_{2{\bf k}}+\bar{\Delta}^{(\nu)4}({\bf k})},~~~~~~~
\end{eqnarray*}
with $b^{(\nu)}_{1{\bf k}}=\varepsilon^{(\nu)2}_{\bf k}
-\varepsilon^{(\nu)2}_{0{\bf k}}+2\bar{\Delta}^{(\nu)2}({\bf k})$, and
$b^{(\nu)}_{2{\bf k}} =(\varepsilon^{(\nu)}_{\bf k}-\varepsilon^{(\nu)}_{0{\bf k}})^{2}
+\bar{\Delta}^{(\nu)2}({\bf k})$, while the coherence factors in the SC-state with the coexisting
electronic order can be obtained as,
\begin{subequations}\label{coherence-factors}
\begin{eqnarray}
U^{(\nu)2}_{1{\bf k}}&=&{1\over 2}\left [a^{(\nu)}_{1{\bf k}}\left (1+{\varepsilon^{(\nu)}_{\bf k}\over
E^{(\nu)}_{1{\bf k}}}\right )- a^{(\nu)}_{3{\bf k}}\left (1+{\varepsilon^{(\nu)}_{0{\bf k}}\over
E^{(\nu)}_{1{\bf k}}}\right )\right ],\\
V^{(\nu)2}_{1{\bf k}}&=&{1\over 2}\left [a^{(\nu)}_{1{\bf k}}\left (1-{\varepsilon^{(\nu)}_{\bf k}\over
E^{(\nu)}_{1{\bf k}}}\right )- a^{(\nu)}_{3{\bf k}}\left (1-{\varepsilon^{(\nu)}_{0{\bf k}}\over
E^{(\nu)}_{1{\bf k}}}\right )\right ],\\
U^{(\nu)2}_{2{\bf k}}&=&-{1\over 2}\left [a^{(\nu)}_{2{\bf k}}\left (1+{\varepsilon^{(\nu)}_{\bf k}\over
E^{(\nu)}_{2{\bf k}}}\right )- a^{(\nu)}_{3{\bf k}}\left (1+{\varepsilon^{(\nu)}_{0{\bf k} }\over
E^{(\nu)}_{2{\bf k}}}\right )\right ],\\
V^{(\nu)2}_{2{\bf k}}&=&-{1\over 2}\left [a^{(\nu)}_{2{\bf k}}\left (1-{\varepsilon^{(\nu)}_{\bf k}\over
E^{(\nu)}_{2{\bf k}}}\right )- a^{(\nu)}_{3{\bf k}}\left (1-{\varepsilon^{(\nu)}_{0{\bf k} }\over
E^{(\nu)}_{2{\bf k}}}\right )\right ], ~~~~~~~
\end{eqnarray}
\end{subequations}
with $a^{(\nu)}_{3{\bf k}}=[\bar{\Delta}^{(\nu)}_{\rm co}({\bf k})]^{2}/(E^{(\nu)2}_{1{\bf k}}
-E^{(\nu)2}_{2{\bf k}})$. These coherence factors are constrained by the sum rule for any wave vector
${\bf k}$, i.e., $U^{(\nu)2}_{1{\bf k}}+V^{(\nu)2}_{1{\bf k}}+ U^{(\nu)2}_{2{\bf k}}+ V^{(\nu)2}_{2{\bf k}}=1$.

From the above expression of the electron normal and anomalous Green's function in Eq. (\ref{EGF1}),
it is easy to find that on the bonding (antibonding) EFS contour
${\bf k} ^{(\rm B)}_{\rm F}$ (${\bf k}^{(\rm A)}_{\rm F}$) as shown in Fig. \ref{spectral-maps}, the bonding
(antibonding) energy eigenvalue $E^{(1)}_{1\bf k}$ ($E^{(2)}_{1\bf k}$) along ${\bf k}^{(\rm B)}_{\rm F}$
(${\bf k}^{(\rm A)}_{\rm F}$) vanishes, while on the bonding (antibonding) sheet ${\bf k}^{(\rm B)}_{\rm BS}$
(${\bf k}^{(\rm A)}_{\rm BS}$), the bonding (antibonding) energy eigenvalue $E^{(1)}_{2\bf k}$
($E^{(2)}_{2\bf k}$) along ${\bf k}^{(\rm B)}_{\rm BS}$ (${\bf k}^{(\rm A)}_{\rm BS}$) is equal to zero.
These results therefore reflect a fact that in analogy to the single-layer case \cite{Zhao17,Gao18,,Gao18a},
the bonding (antibonding) energy band splitting is induced by the bonding (antibonding) electronic order gap,
which leads to the bonding (antibonding) sheet ${\bf k}^{\rm (B)}_{\rm BS}$ (${\bf k}^{\rm (A)}_{\rm BS}$) is
bifurcated from the bonding (antibonding) EFS contour ${\bf k}^{\rm (B)}_{\rm F}$ (${\bf k}^{\rm (A)}_{\rm F}$)
in momentum space except for at around the hot spots ${\bf k}^{\rm (B)}_{\rm HS}$ (${\bf k}^{\rm (A)}_{\rm HS}$)
as shown in Fig. \ref{spectral-maps}. Moreover, both the bonding and antibonding electron self-energies
$\Sigma^{(1)}_{\rm ph}({\bf k},\omega)$ and $\Sigma^{(2)}_{\rm ph}({\bf k},\omega)$ originated from the
electron's coupling to spin excitations are strongly momentum dependent, indicating that both the bonding
and antibonding electronic order gaps $\bar{\Delta}^{(1)}_{\rm co}({\bf k})$ and
$\bar{\Delta}^{(2)}_{\rm co}({\bf k})$ are also strongly dependent on momentum.

\begin{figure}[h!]
\centering
\includegraphics[scale=0.85]{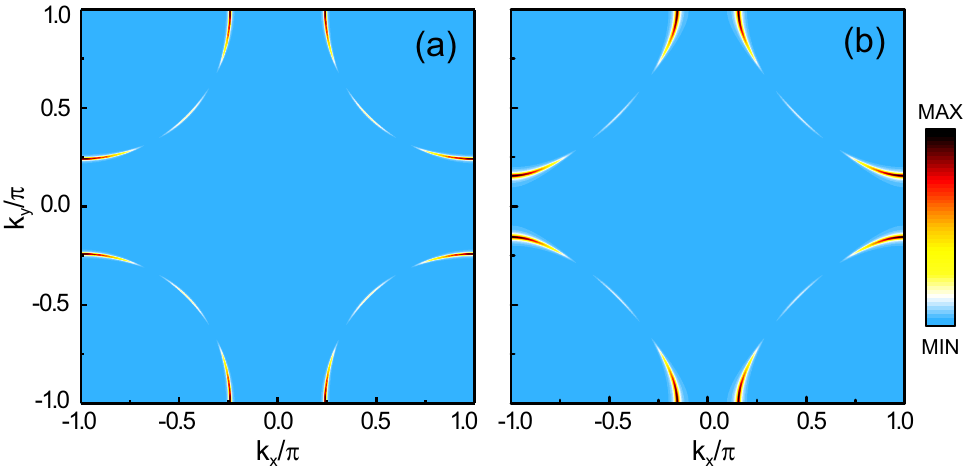}
\caption{(Color online) The intensity map of (a) the bonding and (b) the antibonding quasiparticle scattering
rates at $\delta=0.12$ with $T=0.002J$ for $t/J=2.5$, $t'/t=0.28$, and $t_{\perp}/t=0.3$. \label{scattering-rate}}
\end{figure}

\begin{figure}[t!]
\centering
\includegraphics[scale=0.65]{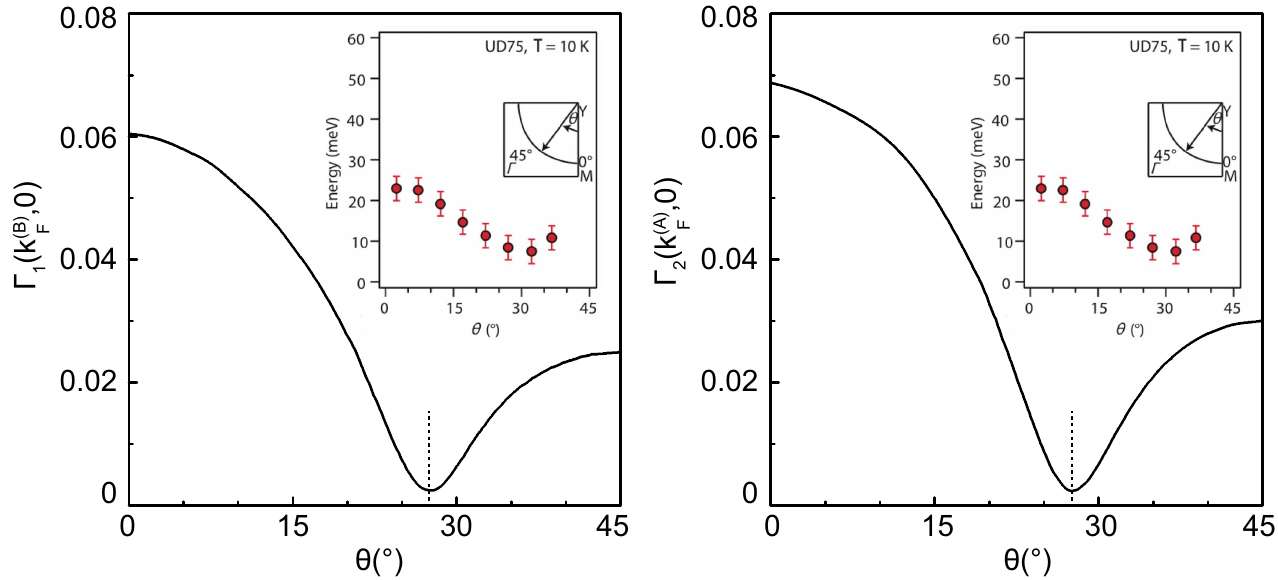}
\caption{(Color online) The angular dependence of (a) the bonding quasiparticle scattering rate along
${\bf k}^{\rm (B)}_{\rm F}$ from the bonding antinode to the node and (b) the antibonding quasiparticle
scattering rate along ${\bf k}^{\rm (A)}_{\rm F}$ from the antibonding antinode to the node at $\delta=0.12$
with $T=0.002J$ for $t/J=2.5$, $t'/t=0.28$, and $t_{\perp}/t=0.3$. The position of the corresponding hot spot
is indicated by the dashed vertical line. Inset in (a) and (b): the experimental result of the angular dependence
of the quasiparticle scattering rate for the underdoped bilayer cuprate superconductors
Bi$_{2}$Sr$_{2}$CaCu$_{2}$O$_{8+\delta}$ taken from Ref. \onlinecite{Vishik09}. \label{AD-scattering-rate}}
\end{figure}

The quasiparticle scattering rate $\Gamma_{\nu}({\bf k},\omega)$ in Eq. (\ref{EQDSR}) is dominated by the
imaginary part of the electron self-energy ${\rm Im}\Sigma^{(\nu)}_{\rm ph}({\bf k},\omega)$ [then the electronic
order gap $\bar{\Delta}^{(\nu)}_{\rm co}({\bf k})$]. As the results shown in Fig. \ref{spectral-maps} and
Fig. \ref{spectral-maps-3D}, the spectral weights in both the bonding and antibonding sheets
${\bf k}^{\rm (B)}_{\rm BS}$ and ${\bf k}^{\rm (A)}_{\rm BS}$ are suppressed, leading to that these two sheets
become unobservable in experiments. However, in order to understand how the renormalization from the quasiparticle
scattering reduces almost all spectral weights in the bonding (antibonding) EFS contour ${\bf k}^{\rm (B)}_{\rm F}$
(${\bf k}^{\rm (A)}_{\rm F}$) to the bonding (antibonding) hot spots, we plot the intensity map of (a) the bonding
quasiparticle scattering rate $\Gamma_{1}({\bf k},0)$ and (b) the antibonding quasiparticle scattering rate
$\Gamma_{2}({\bf k},0)$ at $\delta=0.12$ with $T=0.002J$ in Fig. \ref{scattering-rate}, where the quasiparticle
scattering occurs mainly along the bonding and antibonding EFS contours. However, both $\Gamma_{1}({\bf k},0)$
and $\Gamma_{2}({\bf k},0)$, with the scattering much stronger at around the antinodal region than the nodal region,
are strongly dependent on the Fermi angle. To see these anisotropic quasiparticle scattering rates along the
bonding and antibonding EFS contours more clearly, we plot the angular dependence of (a)
$\Gamma_{1}({\bf k}^{\rm (B)}_{\rm F},0)$ along the bonding EFS contour ${\bf k}^{\rm (B)}_{\rm F}$ and (b)
$\Gamma_{2}({\bf k}^{\rm (A)}_{\rm F},0)$ along the antibonding EFS contour ${\bf k}^{\rm (A)}_{\rm F}$ from the
corresponding antinode to the node at $\delta=0.12$ with $T=0.002J$  in Fig. \ref{AD-scattering-rate} in comparison
with the experimental result \cite{Vishik09} of the quasiparticle scattering rate along EFS contour observed from
the underdoped bilayer cuprate superconductors Bi$_{2}$Sr$_{2}$CaCu$_{2}$O$_{8+\delta}$ (insets). It is thus shown
that both $\Gamma_{1}({\bf k}^{\rm (B)}_{\rm F},0)$ and $\Gamma_{2}({\bf k}^{\rm (A)}_{\rm F},0)$ have a strong
angular dependence, with $\Gamma_{1}({\bf k}^{\rm (B)}_{\rm F}, 0)$ and $\Gamma_{2}({\bf k}^{\rm (A)}_{\rm F},0)$
that exhibit the strongest scattering at the corresponding bonding and antibonding antinodes
${\bf k}^{\rm (B)}_{\rm AN}$ and ${\bf k}^{\rm (A)}_{\rm AN}$, respectively. More interestingly,
$\Gamma_{1}({\bf k}^{\rm (B)}_{\rm F},0)$ and $\Gamma_{2}({\bf k}^{\rm (A)}_{\rm F},0)$ exhibit the weakest
scattering at the corresponding bonding and antibonding hot spots ${\bf k}^{\rm (B)}_{\rm HS}$ and
${\bf k}^{\rm (A)}_{\rm HS}$, respectively. This angular dependences of $\Gamma_{1}({\bf k}^{\rm (B)}_{\rm F},0)$
[$\Gamma_{2}({\bf k}^{\rm (A)}_{\rm F},0)$] therefore suppresses heavily the spectral weight of the electron
quasiparticle excitation spectrum on the bonding (antibonding) EFS contour at around the bonding (antibonding)
antinodes \cite{Loret18,Gao-Q19,Ai19,Chatterjee06,Sassa11,Kaminski05}. However, it suppresses modestly the spectral
weight on the bonding (antibonding) EFS contour at around the nodes \cite{Graf11}, and has the weakest scattering
at around the bonding (antibonding) hot spots. This special momentum dependence of
$\Gamma_{1}({\bf k}^{\rm (B)}_{\rm F},0)$ [$\Gamma_{2}({\bf k}^{\rm (A)}_{\rm F},0)$] therefore reduces almost all
spectral weights on the bonding (antibonding) EFS contour to the bonding (antibonding) hot spots. These hot spots
connected by the scattering wave vectors ${\bf q}_{i}$ construct an {\it octet} scattering model. Concomitantly,
this instability of EFS thus drives the formation of the multiple electronic orders with the wave vectors
${\bf q}_{i}$ connecting the hot spots. As in the single-layer case \cite{Gao18}, these multiple electronic orders
coexist and compete with superconductivity in the bilayer cuprate superconductors. On the other hand, the positions
of the bonding and antibonding hot spots in the bilayer cuprate superconductors are doping dependent. For a better
understanding of the evolution of the positions of the bonding and antibonding hot spots with doping, we have made
a series of calculations for the bonding and antibonding components of the electron quasiparticle excitation spectrum
$I_{1}({\bf k},\omega)$ and $I_{2}({\bf k},\omega)$ with different doping concentrations, and the result shows that
when doping is increased, the positions of the bonding and antibonding hot spots move towards to the corresponding
bonding and antibonding antinodes, respectively, which thus induce a decrease of the charge-order wave vector
connecting the parallel hot spots with the increase of doping
\cite{Comin16,Vishik18,Comin14,Wu11,Chang12,Ghiringhelli12,Neto14,Fujita14,Campi15,Comin15a,Peng16,Hinton16}.
In the normal-state, where the electron pair gap $\bar{\Delta}^{(\nu)}({\bf k},\omega)=0$, the quasiparticle
scattering rate $\Gamma_{\nu}({\bf k},\omega)$ in Eq. (\ref{EQDSR}) is reduced to the normal-state quasiparticle
scattering rate $\Gamma_{\nu}({\bf k},\omega)=|{\rm Im}\Sigma^{(\nu)}_{\rm ph}({\bf k},\omega)|$. However, the
main behavior of the angular dependence of $\Gamma_{\nu}({\bf k},0)$ in the SC-state is dominated by the imaginary
part of the electron self-energy ${\rm Im}\Sigma^{(\nu)}_{\rm ph}({\bf k},\omega)$ [then the electronic order gap
$\bar{\Delta}^{(\nu)}_{\rm co}({\bf k})$]. This is why the main feature of the above {\it octet} scattering model
and the related multiple electronic orders driven by EFS instability can persist into the normal-state
\cite{Chatterjee06,Gao19}.

\subsection{Line-shape in electron quasiparticle excitation spectrum}

\begin{figure*}[t!]
\centering
\includegraphics[scale=0.85]{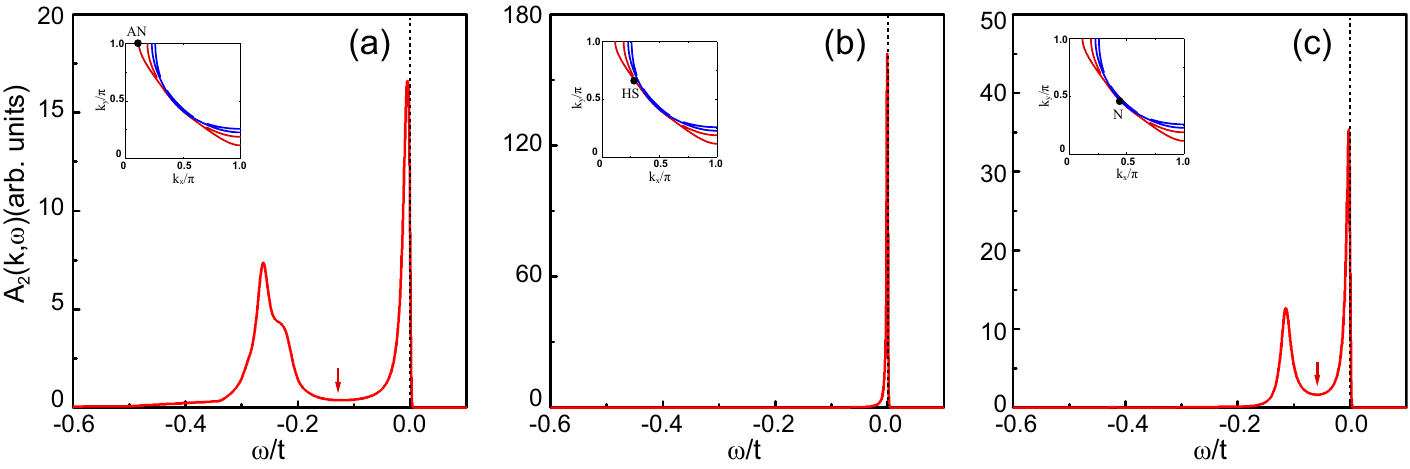}
\caption{(Color online) The antibonding component of the electron quasiparticle excitation spectrum as a
function of energy at (a) the antibonding antinode, (b) the antibonding hot spot, and (c) the node in
$\delta=0.12$ with $T=0.002J$ for $t/J=2.5$, $t'/t=0.28$, and $t_{\perp}/t=0.3$, where the arrows indicate
the positions of the dip, while AN, HS, and N in the insets denote the antibonding antinode, antibonding
hot spot, and node, respectively. \label{PDH-AB}}
\end{figure*}

\begin{figure*}[t!]
\centering
\includegraphics[scale=0.85]{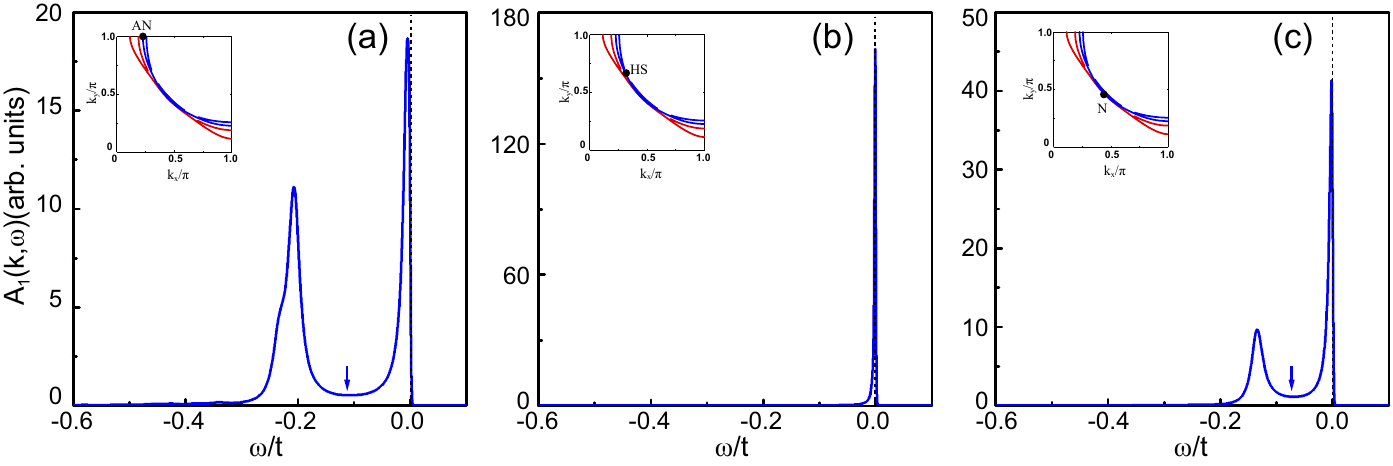}
\caption{(Color online) The bonding component of the electron quasiparticle excitation spectrum as a
function of energy at (a) the bonding antinode, (b) the bonding hot spot, and (c) the node in $\delta=0.12$
with $T=0.002J$ for $t/J=2.5$, $t'/t=0.28$, and $t_{\perp}/t=0.3$, where the arrows indicate the positions
of the dip, while AN, HS, and N in the insets denote the bonding antinode, bonding hot spot, and node,
respectively. \label{PDH-BB}}
\end{figure*}

We now turn to discuss the complicated line-shape in the electron quasiparticle excitation spectrum of the
bilayer cuprate cuprate superconductors. One of the most characteristic features of the electron quasiparticle
excitation spectrum has been the so-called PDH structure observed firstly on the bilayer cuprate superconductors Bi$_{2}$Sr$_{2}$CaCu$_{2}$O$_{8+\delta}$ at around the antinodal region
\cite{Dessau91,Hwu91,Randeria95,Saini97,Norman97,Fedorov99,Campuzano99}. This PDH structure consists of a
sharp electron quasiparticle excitation peak at the lowest binding-energy, a broad hump at the higher
binding-energy, and a spectral dip between them
\cite{Dessau91,Hwu91,Randeria95,Saini97,Norman97,Fedorov99,Campuzano99}. Later, this outstanding PDH structure
was also found in YBa$_{2}$Cu$_{3}$O$_{7-\delta}$ \cite{Lu01} and in other families of cuprate superconductors \cite{Sato02,DLFeng02,Borisenko03,Wei08}. This well-known PDH structure now has been identified along the
entire EFS \cite{Sakai13,Loret17}, and is a hallmark of the spectral line-shape of the ARPES spectrum in cuprate
superconductors. Considering the {\it dip} as a consequence of very strong scattering of the electron
quasiparticles by {\it bosonic mode} immediately nominates the appropriate {\it bosonic excitations} for the
role of the electron pairing glue. Several scenarios have been proposed: the strong electron-phonon coupling,
the bilayer splitting effect, and the pseudogap effect \cite{Eschrig06,Loret18,Kordyuk02,Hashimoto15,DMou17}.
However, the origin of this PDH structure in the electron quasiparticle excitation spectrum is still under debate.

Although the renormalization of the electrons in cuprate superconductors is characterized by a strong
momentum-dependent anisotropy between the electron quasiparticle excitations along EFS, the information revealed
by ARPES experiments has shown that at around the antinodal, the hot spot, and the nodal regions of EFS contain
the essentials of the whole low-energy electron quasiparticle excitations of cuprate superconductors \cite{Damascelli03,Campuzano04,Fink07,Kordyuk14,Zhou18}. In this case, we have made a series of calculations for
the bonding (antibonding) component of the electron quasiparticle excitation spectrum $I_{1}({\bf k},\omega)$
[$I_{2}({\bf k},\omega)$] along with the bonding (antibonding) EFS ${\bf k}^{\rm (B)}_{\rm F}$
(${\bf k}^{\rm (A)}_{\rm F}$) from the bonding (antibonding) antinode to the node. In Fig. \ref{PDH-AB}, we
firstly plot the results of the antibonding component of the electron quasiparticle excitation spectrum
$I_{2}({\bf k}, \omega)$ as a function of energy at (a) the antibonding antinode, (b) the antibonding hot spot,
and (c) the node for $\delta=0.12$ with $T=0.002J$. At around the antibonding antinode (see Fig. \ref{PDH-AB}a),
a very sharp antibonding peak emerges at the lowest binding-energy corresponding to the SC peak, followed by an
antibonding dip and then an antibonding hump in the higher binding-energy, leading to form a PDH structure in the
antibonding component of the electron quasiparticle excitation spectrum
\cite{Dessau91,Hwu91,Randeria95,Saini97,Norman97,Fedorov99,Campuzano99,Lu01}. However, the positions of the
antibonding peak, the antibonding dip, and the antibonding hump are momentum dependent. In particular, the
position of the antibonding hump moves appreciably towards the antibonding peak as the momentum moves from the
antibonding antinode to the antibonding hot spot, and this antibonding hump is eventually incorporated with the
antibonding peak at around the antibonding hot spot (see Fig. \ref{PDH-AB}b), which leads to an absence of the
PDH structure at around the antibonding hot spot. More importantly, this PDH structure is gradually developed
again as the momentum moves from the antibonding hot spot to the node (see Fig. \ref{PDH-AB}c). This evolution
of the PDH structure with momentum in the antibonding component of the electron quasiparticle excitation spectrum
in the bilayer cuprate superconductors is in a striking analogy to that in the quasiparticle excitation spectrum
of the pure single-layer case \cite{Zhao17,Gao18,Gao18a}.

As a complement of the above analysis of the spectral line-shape in the antibonding component of the electron
quasiparticle excitation spectrum, the results of the bonding component of the electron quasiparticle excitation
spectrum $I_{1}({\bf k},\omega)$ as a function of energy at (a) the bondig antinode, (b) the bonding hot spot,
and (c) the node in $\delta=0.12$ with $T=0.002J$ are plotted in Fig. \ref{PDH-BB}. It is remarkable that the
evolution of the PDH structure with momentum in the bonding component of the electron quasiparticle excitation
spectrum is almost the same as that in the antibonding component of the electron quasiparticle excitation spectrum,
including the location of the dip energy. These results in Fig. \ref{PDH-AB} and Fig. \ref{PDH-BB} confirm again
that the antibonding component of the electron quasiparticle excitation spectrum in the bilayer cuprate
superconductors is completely independent of the bonding component.

\begin{figure}[t!]
\centering
\includegraphics[scale=1.0]{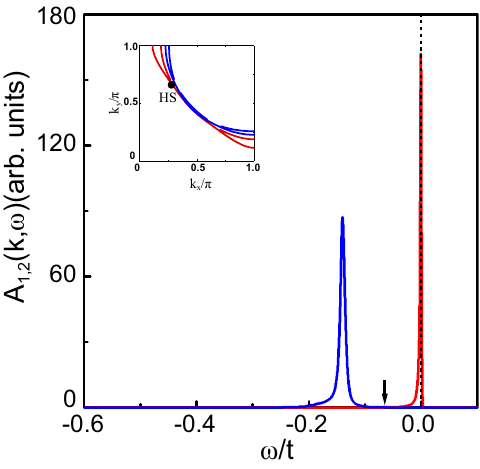}
\caption{(Color online) The bonding (blue line) and antibonding (red line) components of the electron
quasiparticle excitation spectrum as a function of energy at the antibonding hot spot in $\delta=0.12$ with
$T=0.002J$ for $t/J=2.5$, $t'/t=0.28$, and $t_{\perp}/t=0.3$, where the arrow indicates the position of the
dip, while HS in the insets denotes the antibonding hot spot. \label{PDH-HS}}
\end{figure}

\begin{figure*}[t!]
\centering
\includegraphics[scale=0.9]{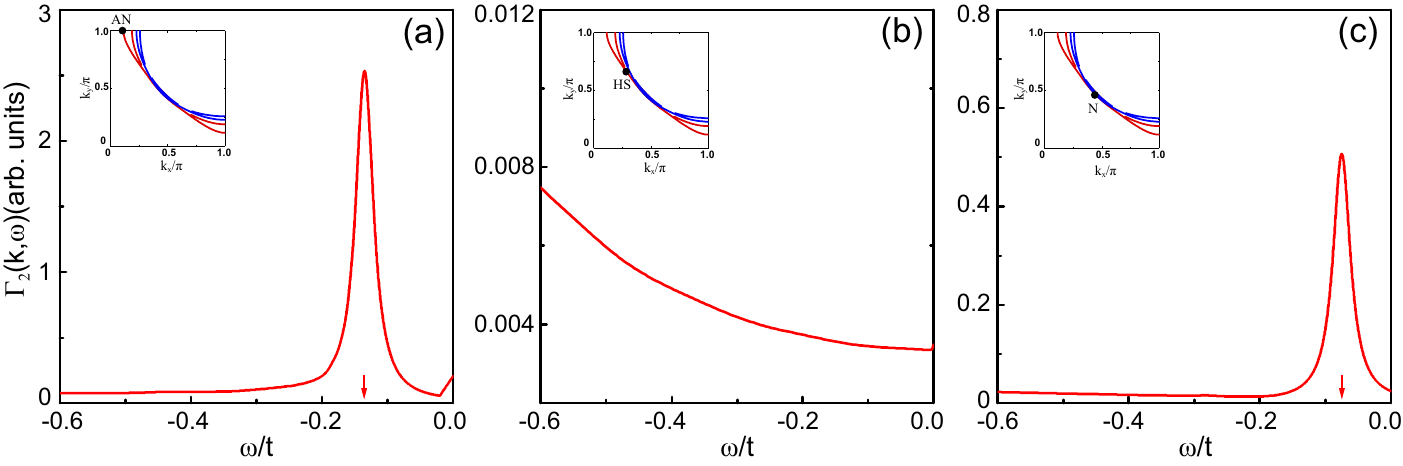}
\caption{(Color online) The antibonding quasiparticle scattering rate as a function of energy at (a) the
antibonding antinode, (b) the antibonding hot spot, and (c) the node as a function of energy at
$\delta=0.12$ with $T=0.002J$ for $t/J=2.5$, $t'/t=0.28$, and $t_{\perp}/t=0.3$, where the red arrows
indicate the positions of the peaks, while AN, HS, and N in the insets denote the antibonding antinode,
antibonding hot spot, and node, respectively. \label{scattering-rate-AB}}
\end{figure*}

Although the PDH structure in the bonding (antibonding) component of the electron quasiparticle excitation
spectrum vanishes at around the bonding (antibonding) hot spots, the contribution from both the bonding and
antibonding components of the electron quasiparticle excitation spectrum generates the PDH structure at around
the antibonding hot spots. To see this special feature more clearly, we plot the result of the bonding and
antibonding components of the quasiparticle excitation spectrum $I_{1}({\bf k},\omega)$ (blue line) and
$I_{2}({\bf k},\omega)$ (red line) as a function of energy at the antibonding hot spot for $\delta=0.12$ with
$T=0.002J$ in Fig. \ref{PDH-HS}. The result in Fig. \ref{PDH-HS} therefore shows clearly that the PDH structure
at around the antibonding hot spot is mainly caused by the pure bilayer coupling
\cite{Kordyuk02,Kordyuk06a,Kordyuk10}, with the very sharp low-energy peak that is closely related to the
antibonding component of the quasiparticle excitation spectrum, while the hump is directly formed by the
bonding component. In particular, this PDH structure at the antibonding hot spots originates from the pure
bilayer coupling at any doping levels \cite{Kordyuk02,Kordyuk06a,Kordyuk10}, which is in sharp contrast to the
pure single-layer cuprate superconductors \cite{Zhao17,Gao18,Gao18a}, where the PDH structure is absent from
the hot spots at any doping levels. Our results in Fig. \ref{PDH-AB}, Fig. \ref{PDH-BB}, and Fig. \ref{PDH-HS}
thus show that the PDH structure in the bilayer cuprate superconductors is significantly modulated by the
hybridization of two copper-oxide layers within a unit cell, resulting in the PDH structure that can be
present all around EFS.

In summary, we have found the following results within the framework of the kinetic-energy driven
superconductivity: (i) the electron quasiparticle excitation spectrum in the bilayer cuprate superconductors
is split into its bonding and antibonding components, with each component itself that contributes to the
low-energy spectral line-shape; (ii) although the PDH structure in the bonding (antibonding) component of the
electron quasiparticle excitation spectrum is absent from the bonding (antibonding) hot spots, the total
contribution from both the bonding and antibonding components of the electron quasiparticle excitation spectrum
generates the PDH structure at around the antibonding hot spots, with the sharp low-energy peak that is directly
associated with the antibonding component, while the hump is mainly dominated by the bonding component.
These results of the PDH structure in the electron quasiparticle excitation spectrum are well consistent with
the experimental observations on the bilayer cuprate superconductors
\cite{Dessau91,Hwu91,Randeria95,Saini97,Norman97,Fedorov99,Campuzano99,Lu01,Sakai13,Loret17,DMou17}.

The bonding (antibonding) component of the electron quasiparticle excitation spectrum in
Eq. (\ref{QPE-spectrum}) [then the electron spectral function in Eq. (\ref{ESF})] has a well-defined resonance
character, where $I_{\nu}({\bf k}, \omega)$ exhibits the peaks when the incoming photon energy $\omega$ is equal
to the renormalized quasiparticle excitation energy, i.e.,
\begin{eqnarray*}\label{band}
\omega -\bar{E}_{\nu}({\bf k},\omega)=0,
\end{eqnarray*}
and then the weights of these peaks are dominated by the inverse of the quasiparticle scattering rate
$\Gamma_{\nu}({\bf k},\omega)$. In other words, the spectral line-shape in the electron quasiparticle
excitation spectrum is determined by both the renormalized band structure $\bar{E}_{\nu}({\bf k}, \omega)$
and the quasiparticle scattering rate $\Gamma_{\nu}({\bf k},\omega)$.

\begin{figure*}[t!]
\centering
\includegraphics[scale=0.9]{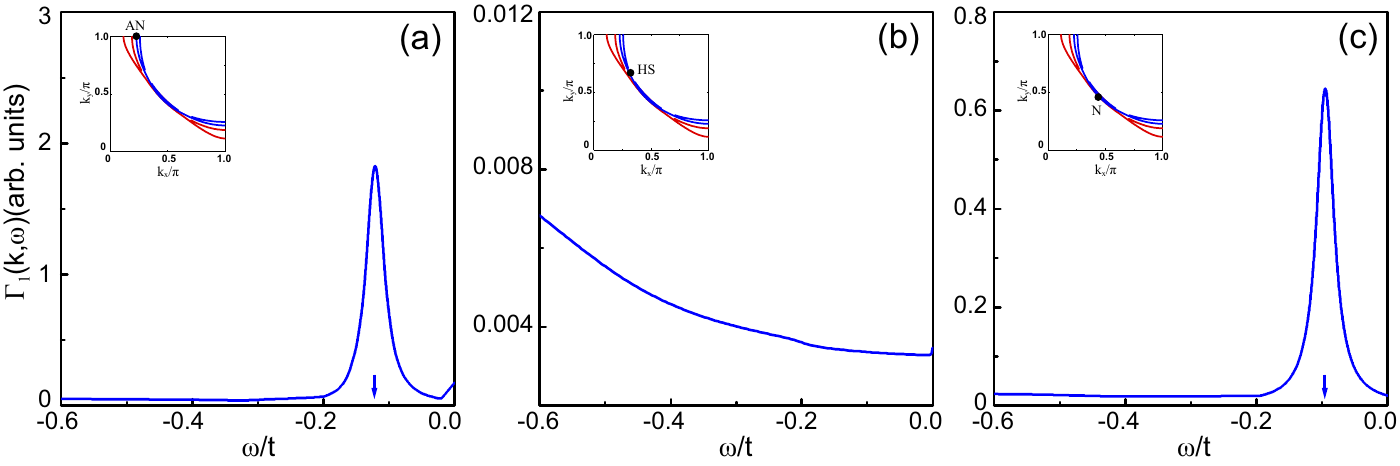}
\caption{(Color online) The bonding quasiparticle scattering rate as a function of energy at (a) the
bonding antinode, (b) the bonding hot spot, and (c) the node as a function of energy at $\delta=0.12$
with $T=0.002J$ for $t/J=2.5$, $t'/t=0.28$, and $t_{\perp}/t=0.3$, where the blue arrows indicate the
positions of the peaks, while AN, HS, and N in the insets denote the bonding antinode, bonding hot spot,
and node, respectively. \label{scattering-rate-BB}}
\end{figure*}

In the electron quasiparticle excitation spectrum, a quasiparticle with a long lifetime is observed as
a sharp peak in intensity, and a quasiparticle with a short lifetime is observed as a broad hump, while
the dip in the electron quasiparticle excitation spectrum except for at around the bonding (antibonding)
hot spots implies that the quasiparticle scattering rate $\Gamma_{\nu}({\bf k},\omega)$ has a peak structure
with the sharp peak located at around the dip energy. To see this point more clearly, we first plot the
antibonding quasiparticle scattering rate $\Gamma_{2}({\bf k}, \omega)$ as a function of energy at (a) the
antibonding antinode, (b) the antibonding hot spot, and (c) the node for $\delta=0.12$ with $T=0.002J$ in
Fig. \ref{scattering-rate-AB}. At around the antibonding antinode, the antibonding quasiparticle scattering
rate $\Gamma_{2}({\bf k},\omega)$ consists of a peak structure, where $\Gamma_{2}({\bf k},\omega)$ reaches a
sharp peak at the binding-energy of $\omega_{\rm AB}\sim 0.136t$, and then the weight of this sharp peak
decreases rapidly away from this binding-energy $\omega_{\rm AB}$, in good agreement with the corresponding
experimental result \cite{DMou17} of the energy dependence of the quasiparticle scattering rate at around
the antinodal region observed on the bilayer cuprate superconductors Bi$_{2}$Sr$_{2}$CaCu$_{2}$O$_{8+\delta}$.
In particular, the position of this sharp peak in $\Gamma_{2}({\bf k}, \omega)$ is just corresponding to the
position of the antibonding dip in the antibonding component of the electron quasiparticle excitation spectrum
as shown in Fig. \ref{PDH-AB}a. In other words, the peak structure in $\Gamma_{2}({\bf k},\omega)$ induces an
intensity depletion in the antibonding dip region. However, the peak in $\Gamma_{2}({\bf k},\omega)$ at around
the antinode becomes progressively broader as the antibonding hot spot is approached, and then the peak
structures in $\Gamma_{2}({\bf k},\omega)$ vanish eventually at the antibonding hot spot as shown in
Fig. \ref{scattering-rate-AB}b, which therefore leads to that the PDH structure in the antibonding component
of the quasiparticle excitation spectrum is absent from the antibonding hot spot as shown in
Fig. \ref{PDH-AB}b. Moreover, the peak in $\Gamma_{2}({\bf k},\omega)$ is gradually developed again as the
momentum moves away from the antibonding hot spot to the node (see Fig. \ref{scattering-rate-AB}c), which
induces the emergence of the PDH structure at around the nodal region (see Fig. \ref{PDH-AB}c).

In Fig. \ref{scattering-rate-BB}, we plot the bonding quasiparticle scattering rate
$\Gamma_{1}({\bf k},\omega)$ as a function of energy at (a) the bonding antinode, (b) the bonding hot spot,
and (c) the node for $\delta=0.12$ with $T=0.002J$. Obviously, the evolution of the bonding quasiparticle
scattering rate $\Gamma_{1}({\bf k},\omega)$ with momentum in Fig. \ref{scattering-rate-BB} is the same as
that of the antibonding quasiparticle scattering rate $\Gamma_{2}({\bf k},\omega)$ in
Fig. \ref{scattering-rate-AB}. This special momentum dependence of the peak structure in the bonding
quasiparticle scattering rate therefore induces the emergence of the striking PDH structure in the bonding
component of the quasiparticle excitation spectrum shown in Fig. \ref{PDH-BB}.

\subsection{Kink in dispersion}

\begin{figure*}[t!]
\centering
\includegraphics[scale=0.8]{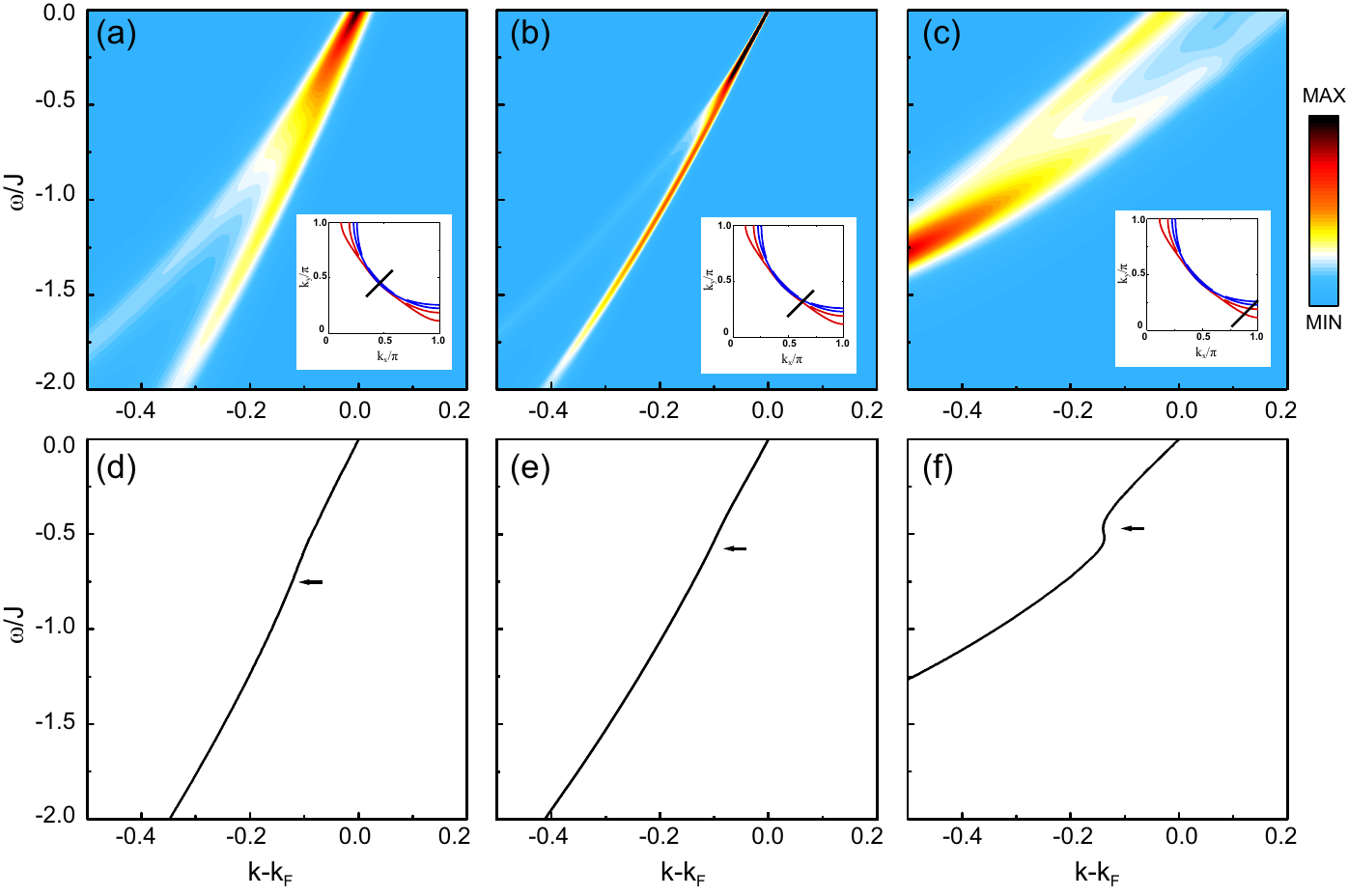}
\caption{(Color online) Upper panel: the intensity maps of the antibonding component of the electron
quasiparticle excitation spectrum as a function of binding-energy along (a) the nodal cut, (b) the
hot spot cut, and (c) the antinodal cut at $\delta=0.12$ with $T=0.002J$ for $t/J=2.5$, $t'/t=0.28$,
and $t_{\perp}/t=0.3$. Lower panel: the antibonding electron quasiparticle dispersions along (d) the
nodal cut, (e) the hot spot cut, and (f) the antinodal cut extracted from the positions of the
lowest-energy electron quasiparticle excitation peaks in (a), (b) and (c), respectively. The arrow
indicates the kink position. \label{kink-maps}}
\end{figure*}

The first indication of the renormalization of the electrons in cuprate superconductors was observed
in the ARPES measurements on the bilayer cuprate superconductor Bi$_{2}$Sr$_{2}$CaCu$_{2}$O$_{8+\delta}$,
where the presence of the kink in the electron quasiparticle dispersion is identified along the nodal direction
\cite{Bogdanov00,Lanzara01,Cuk04,Iwasawa08,Anzai10,He18,Yang19,Kaminski01,Johnson01,Sato03,Kordyuk06,Plumb13,He13}.
Later, this kink is found to be present all around EFS, and appears in all families of cuprate superconductors,
with the energy scale (in the energy range $50\sim 80$ meV) at which the kink occurs that is similar for these
cuprate superconductors with one or more copper-oxide layers per unit cell \cite{Zhou03,Yoshida07,Lee09,Chen09}.
Although it is believed that the kink in the electron quasiparticle dispersion is due to the coupling of the
electrons to particular bosonic excitations, the nature of these bosonic excitations remains controversial,
where two main proposals are disputing the explanations of the origin of the kink. In one of the proposals, the
kink is associated to the phonon \cite{Bogdanov00,Lanzara01,Cuk04,Iwasawa08,Anzai10,He18,Yang19}, while in the
other, the kink is related to the spin excitation \cite{Kaminski01,Johnson01,Sato03,Kordyuk06,Plumb13}. In this
subsection, we show within the framework of the kinetic-energy driven superconductivity that the electron
quasiparticle dispersion is affected by the spin excitation, and then the kink in the electron quasiparticle
dispersion associated with the renormalization of the electrons is electronic in nature, while the bilayer
coupling leads to the enhancement of the kink effect.

To show how the kink dispersion evolves as the momentum moves along EFS, we plot the intensity map of the
antibonding component of the electron quasiparticle excitation spectrum as a function of binding-energy
along (a) the nodal cut, (b) the hot spot cut, and (c) the antinodal cut at $\delta=0.12$ with $T=0.002J$ in
the upper panel of Fig. \ref{kink-maps}. In the lower panel, we plot the corresponding antibonding dispersions
along (d) the nodal cut, (e) the hot spot cut, and (f) the antinodal cut extracted from the positions of the
lowest-energy electron quasiparticle excitation peaks in (a), (b) and (c), respectively. Apparently, the ARPES
experimental results \cite{Kaminski01} of the dispersion kink for the bilayer cuprate superconductors
Bi$_{2}$Sr$_{2}$CaCu$_{2}$O$_{8+\delta}$ are qualitative reproduced. The kink in the electron quasiparticle
dispersion is present all around the antibonding EFS. In particular, the electron quasiparticle dispersion in
the vicinity of the antibonding EFS along the nodal cut is linear in the low-energy and high-energy ranges,
but with different slopes, and then these two ranges are separated by a kink (see Fig. \ref{kink-maps}a and
Fig. \ref{kink-maps}d), which is in a clear contrast to the result obtained from the band structure calculations,
where the linear dispersion from the low-energy to high-energy is predicted. Moreover, for the binding-energy
less than the kink energy, the spectrum exhibits sharp peaks with a weak dispersion, while beyond this, the
broad peaks with a stronger dispersion (see Fig. \ref{kink-maps}a and Fig. \ref{kink-maps}d). However, with
leaving the node, the dispersion kink becomes more dramatic, and in particular, near the cut close to the
antinode, develops into a break separating of the fasting dispersing high-energy part of the electron
quasiparticle excitation spectrum from the slower dispersing low-energy part (see Fig. \ref{kink-maps}c
and Fig. \ref{kink-maps}f), while this breaking effect becomes the most pronounced along the antinodal cut.
The above results in Fig. \ref{kink-maps} therefore reflect an experimental fact that there is a continuous
evolution along EFS from the kink to the break. Concomitantly, the kinks emerge at around the energy
$\omega_{\rm kink}\sim 0.7J=70$ meV in the nodal direction and $\omega_{\rm kink}\sim 0.5J=50$ meV near the
antinode, also in good agreement with the experimental observations
\cite{Bogdanov00,Lanzara01,Cuk04,Iwasawa08,Anzai10,He18,Yang19,Kaminski01,Johnson01,Sato03,Kordyuk06,Plumb13}.
These results also indicate that the characteristic kink energy decreases smoothly as the momentum moves from
the nodal region to the antinodal region.

\begin{figure*}[t!]
\centering
\includegraphics[scale=0.8]{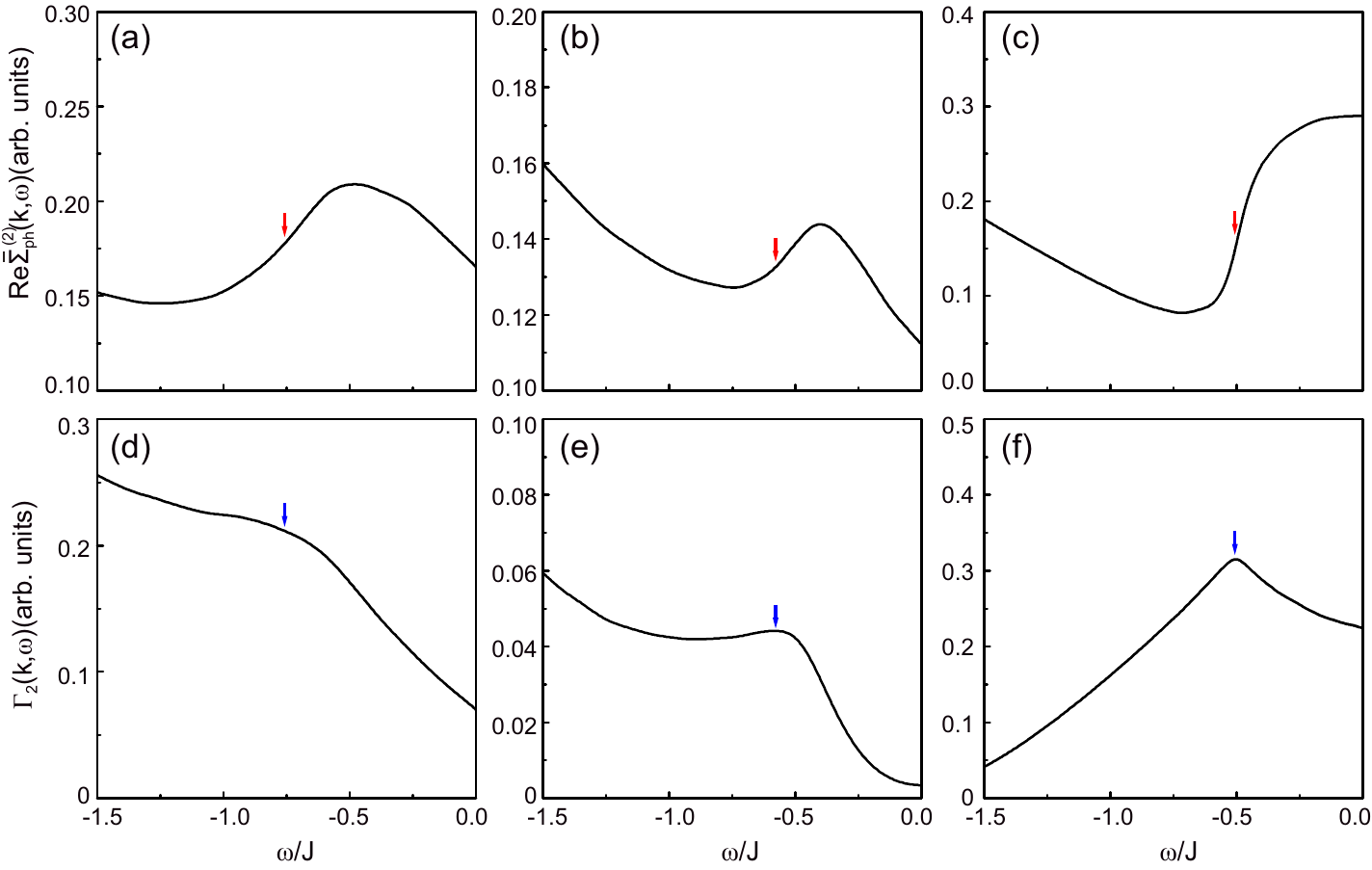}
\caption{(Color online) Upper panel: the real part of the antibonding modified electron self-energy as a
function of binding-energy along (a) the nodal dispersion, (b) the hot spot dispersion, and (c) the
antinodal dispersion at $\delta=0.12$ with $T=0.002J$ for $t/J=2.5$, $t'/t=0.28$, and $t_{\perp}/t=0.3$.
Lower panel: the corresponding antibonding quasiparticle scattering rate as a function of binding-energy
along (d) the nodal dispersion, (e) the hot spot dispersion, and (f) the antinodal dispersion. The red arrow
indicates inflection point, while the blue arrow denotes the peak position. \label{band-structure}}
\end{figure*}

In the framework of the kinetic-energy driven superconductivity, the renormalization of the electrons
originates from the coupling between the electrons and spin excitations. However, the quantity that
describes this coupling is the real and imaginary parts of the electron self-energy
$\Sigma^{(\nu)}_{\rm ph}({\bf k},\omega)$. In this case, above obtained kink in the electron quasiparticle
dispersion relation in Eq. (\ref{MRESE}) that does not originate from the bare dispersion relation
$\varepsilon^{(\nu)}_{\bf k}$ must be due to the slope changes in the real part of the modified electron
self-energy ${\rm Re}{\bar\Sigma}^{(\nu)}_{\rm ph}({\bf k},\omega)$ in Eq. (\ref{RESE}) and the drop in
the quasiparticle scattering rate $\Gamma_{\nu}({\bf k},\omega)$ in Eq. (\ref{EQDSR}). To reveal the deep
relation between the kink and electron self-energy more clearly, we plot the real part of the antibonding
modified electron self-energy ${\rm Re}{\bar\Sigma}^{(2)}_{\rm ph}({\bf k},\omega)$ as a function of
binding-energy along (a) the nodal dispersion, (b) the hot spot dispersion, and (c) the antinodal dispersion
as shown in Fig. \ref{kink-maps}d, Fig. \ref{kink-maps}e, Fig. \ref{kink-maps}f, respectively, at
$\delta=0.12$ with $T=0.002J$ in the upper panel of Fig. \ref{band-structure}, where the red arrow indicates
the inflection point (then the point of the slope change). In the lower panel, we plot the corresponding
antibonding quasiparticle scattering rate $\Gamma_{2}({\bf k},\omega)$ as a function of binding-energy
along (d) the nodal dispersion, (e) the hot spot dispersion, and (f) the antinodal dispersion, where the
blue arrow denotes the peak position (then the point of the drop in the quasiparticle scattering rate). The
results in Fig. \ref{band-structure} therefore show that a slope change in the real part of the modified
electron self-energy ${\rm Re}{\bar\Sigma}^{(2)}_{\rm ph}({\bf k},\omega)$ is present all around the
antibonding EFS, while the corresponding quasiparticle scattering rate $\Gamma_{2}({\bf k},\omega)$ exhibits
a peak structure except for along the nodal direction, where $\Gamma_{2}({\bf k},\omega)$ exhibits a gentle
reduction with the decrease of binding-energy. In this case, we therefore find that the appearance of the
kink in the electron quasiparticle dispersion is always related to the slope change in
${\rm Re}{\bar\Sigma}^{(2)}_{\rm ph} ({\bf k},\omega)$, i.e., the position of the kink in the electron
quasiparticle dispersion shown in Fig. \ref{kink-maps} is exactly the same as that for the corresponding
inflection point in ${\rm Re}{\bar\Sigma}^{(2)}_{\rm ph}({\bf k},\omega)$ shown in Fig. \ref{band-structure}.
In other words, this inflection point is the point of the slope change in
${\rm Re}{\bar\Sigma}^{(2)}_{\rm ph}({\bf k},\omega)$ as shown in Fig. \ref{band-structure}, and therefore
leads to the emergence of the kink in the electron quasiparticle dispersion. This is why the kink in the
electron quasiparticle dispersion marks the crossover between two different slopes. On the other hand, the
position of the kink in the electron quasiparticle dispersion shown in Fig. \ref{kink-maps} is also the same
as that for the corresponding peak in the quasiparticle scattering rate $\Gamma_{2}({\bf k},\omega)$ shown
in Fig. \ref{band-structure} except for along the nodal direction, i.e., there is an exact one to one
correspondence between the kink position shown in Fig. \ref{kink-maps} and the peak position in
$\Gamma_{2}({\bf k},\omega)$ shown in Fig. \ref{band-structure} except for along the nodal cut. This peak
in the quasiparticle scattering rate suppresses heavily the spectral weight at around the kink, and then
the weak spectral intensity appears always at around the kink along the EFS cut except for along the nodal
direction. This is why the experimentally observed kink is always related to the drop in the quasiparticle
scattering rate except for along the nodal cut. More specifically, the scattering along the antinodal
dispersion for the binding-energy less than the kink energy is stronger than that of the corresponding
scattering along the nodal dispersion, while the scattering along the nodal dispersion for the binding-energy
above the kink energy is stronger than that of the corresponding scattering along the antinodal dispersion,
which leads to a fact that the spectral weights along the nodal dispersion above the kink energy and along
the antinodal dispersion less than the kink energy are suppressed heavily.

Although the kink is present all around EFS, there are some subtle differences for the physical origin of the
kinks along the nodal, the hot spot, and the antinodal directions. At the nodal direction, the bilayer coupling
is absent, and then the less visible kink in the antibonding electron quasiparticle dispersion is caused mainly
by the renormalization of the electrons originated from the coupling between the electrons and spin excitations
within a copper-oxide layer. However, the actual minimum of the quasiparticle scattering rate locates exactly at
the hot spot regime as shown in Fig. \ref{AD-scattering-rate}b, where the quasiparticle scattering is very weak,
and then the kink along the hot spot cut is induced mainly by the bilayer coupling. On the other hand, near the
antinode, both the quasiparticle scattering rate and bilayer coupling exhibit the largest value, this large band
splitting together with the particularly strong quasiparticle scattering lead to a break separating of the fasting
dispersing high-energy part of the electron quasiparticle excitation spectrum from the slower dispersing low-energy
part, and then these two branches approach one another at the kink energy. This separation of the band dispersion
becomes most prominent at the antinode and results in the most strong strength of the kink. In other words, although
the kink in the electron quasiparticle dispersion is present all around the antibonding EFS, the kink along the
hot spot cut is different from the kink along the nodal or antinodal cut, while the dispersion kink near the
antinode becomes more pronounced due to the presence of the strong bilayer coupling. The above analysis also explains
why the kink effect near the antinode is more pronounced than that along the nodal cut
\cite{Bogdanov00,Lanzara01,Cuk04,Iwasawa08,Anzai10,He18,Yang19,Kaminski01,Johnson01,Sato03,Kordyuk06,Plumb13}.

As a complement of the above analysis of the kink in the electron quasiparticle dispersion in the SC-state,
we have also studied the dispersion kink in the normal-state, where the electron pair gap
$\bar{\Delta}^{(\nu)}({\bf k},\omega)=0$, and then the real part of the modified electron self-energy
${\rm Re}{\bar\Sigma}^{(\nu)}_{\rm ph}({\bf k},\omega)$ in Eq. (\ref{RESE}) and the quasiparticle scattering
rate $\Gamma_{\nu}({\bf k}, \omega)$ in Eq. (\ref{EQDSR}) are reduced as
${\rm Re}{\bar\Sigma}^{(\nu)}_{\rm ph}({\bf k},\omega)={\rm Re}\Sigma^{(\nu)}_{\rm ph}({\bf k},\omega)$ and
$\Gamma_{\nu}({\bf k},\omega)=|{\rm Im}\Sigma^{(\nu)}_{\rm ph}({\bf k},\omega)|$, respectively. In this case,
we have calculated the antibonding component of the electron quasiparticle excitation spectrum in the
normal-state along the nodal, the hot spot, and the antinodal cuts, and the related real and imaginary parts
of the antibonding electron self-energy ${\rm Re}\Sigma^{(2)}_{\rm ph}({\bf k},\omega)$ and
${\rm Im} \Sigma^{(2)}_{\rm ph}({\bf k},\omega)$ along the nodal, the hot spot, and the antinodal dispersions
with the same set of parameters as in Fig. \ref{kink-maps} and Fig. \ref{band-structure} except for in the
normal-state, and the obtained intensity maps of the antibonding component of the electron quasiparticle
excitation spectrum and the related results of ${\rm Re}\Sigma^{(2)}_{\rm ph}({\bf k},\omega)$ and
${\rm Im}\Sigma^{(2)}_{\rm ph}({\bf k},\omega)$ are almost the same as that of the corresponding maps in
the SC-state in Fig. \ref{kink-maps} and results of ${\rm Re} {\bar\Sigma}^{(2)}_{\rm ph} ({\bf k},\omega)$
and $\Gamma_{2}({\bf k},\omega)$ in Fig. \ref{band-structure}, respectively. These results therefore confirm
that the kink emerged in the SC-state can persist into the normal-state, and is caused by the same electron
self-energy (then the electronic order gap) effect generated by the coupling between the electrons and spin
excitations, also in good agreement with the experimental observations
\cite{Bogdanov00,Lanzara01,Cuk04,Iwasawa08,Anzai10,He18,Yang19,Kaminski01,Johnson01,Sato03,Kordyuk06,Plumb13}.
These results also indicate that the kink in the electron quasiparticle dispersion is totally unrelated to
superconductivity
\cite{Bogdanov00,Lanzara01,Cuk04,Iwasawa08,Anzai10,He18,Yang19,Kaminski01,Johnson01,Sato03,Kordyuk06,Plumb13}.

\subsection{ARPES autocorrelation}\label{ARPES-autocorrelation}

The ARPES autocorrelation
${\bar C}({\bf q},\omega)=(1/N)\sum_{\bf k}I({\bf k}+{\bf q},\omega)I({\bf k},\omega)$ is defined as
the autocorrelation of the electron quasiparticle excitation spectral intensities at two different momenta,
separated by a momentum transfer ${\bf q}$, at a fixed energy $\omega$, where the summation of momentum
${\bf k}$ is restricted within the first BZ \cite{Chatterjee06}. This ARPES autocorrelation is effectively
the momentum-resolved joint density of states in the electronic state, and can give us important new insights
into the renormalization of the electrons. On the other hand, the scanning tunneling spectroscopy (STS)
provides the information on the local density of states (LDOS) of the electronic state as a function of energy,
and although the measured data are obtained in real-space, these data can be inverted in terms of the Fourier
transform (FT) to provide the momentum-space picture of the renormalization of the electrons. This technique
is complementary to the photoemission in that it reveal local variations of the renormalized electrons in
cuprate superconductors \cite{Devereaux07,Fischer07}. In particular, the ARPES experimental results
\cite{Chatterjee06,He14} have shown that the ARPES autocorrelation exhibits discrete spots in momentum-space,
which are directly related with the wave vectors ${\bf q}_{i}$ connecting the hot spots shown in
Fig. \ref{spectral-maps}, and are well consistent with the QSI peaks observed from the FT-STS experiments \cite{Pan01,Hoffman02,Kohsaka07,Kohsaka08,Hamidian16}. In this subsection, we discuss the characteristic
feature of the ARPES autocorrelation in the bilayer cuprate superconductors and its connection with the QSI
measured from the FT-STS experiments.

\begin{figure}[t!]
\centering
\includegraphics[scale=0.75]{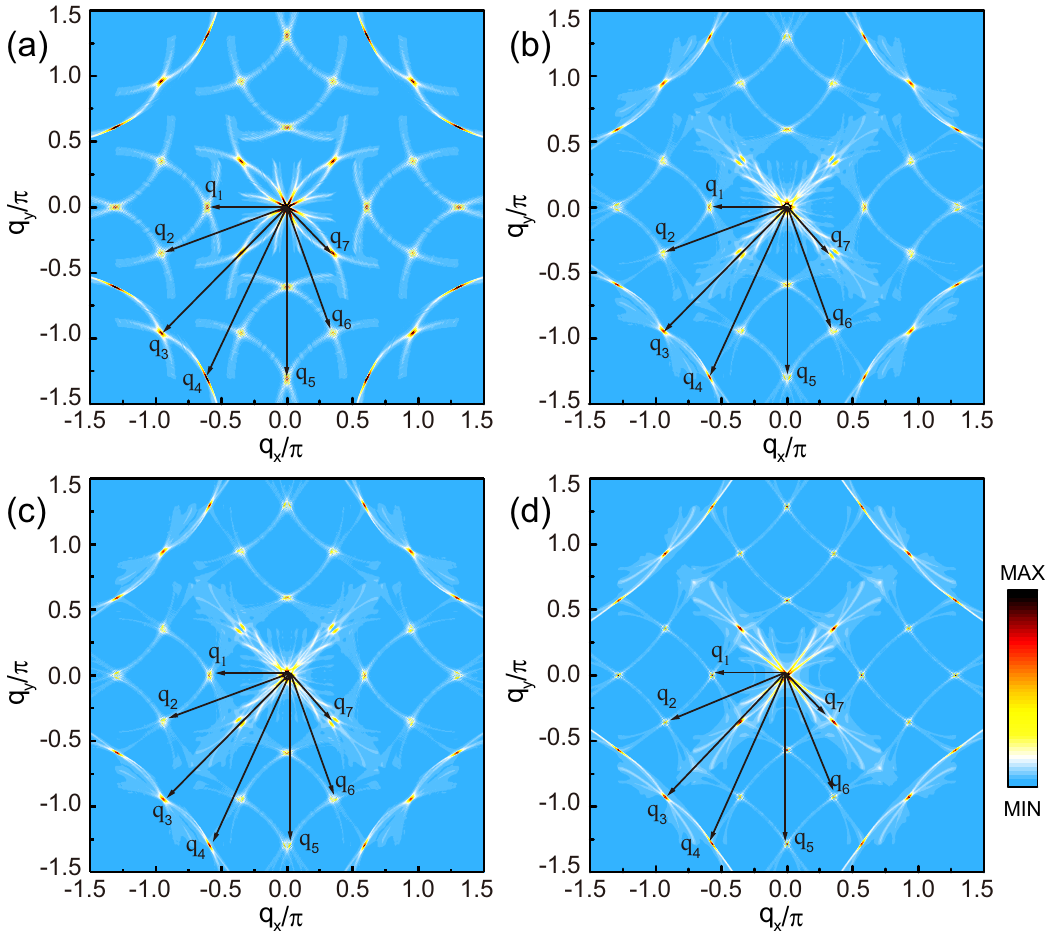}
\caption{(Color online) The intensity map of the autocorrelations of (a) the bonding-bonding spectral
intensities, (b) the bonding-antibonding spectral intensities, (c) the antibonding-bonding spectral
intensities, and (d) the antibonding-antibonding spectral intensities in the $[q_{x},q_{y}]$ plane for
$\omega=0.12J$ at $\delta=0.15$ with $T=0.002J$ for $t/J=2.5$, $t'/t=0.28$, and $t_{\perp}/t=0.3$.
\label{autocorrelation-maps}}
\end{figure}

In the bilayer cuprate superconductors, the experimentally measurable ARPES autocorrelation
${\bar C}({\bf q},\omega)$ can be described in terms of the antibonding and bonding components of the
electron quasiparticle excitation spectrum as,
\begin{subequations}\label{ACF}
\begin{eqnarray}
{\bar C}({\bf q},\omega)&=&\sum_{\nu\nu'}{\bar C}^{(\nu\nu')}({\bf q},\omega), \label{ACF-sum} \\
{\bar C}^{(\nu\nu')}({\bf q},\omega)&=&{1\over N}\sum_{\bf k}I_{\nu}({\bf k}+{\bf q},\omega)
I_{\nu'}({\bf k},\omega), \label{ACF-AB}~~~~~~~
\end{eqnarray}
\end{subequations}
where the components ${\bar C}^{(11)}({\bf q},\omega)$, ${\bar C}^{(12)}({\bf q},\omega)$,
${\bar C}^{(21)}({\bf q},\omega)$, and ${\bar C}^{(22)} ({\bf q},\omega)$ are the autocorrelation
functions of the bonding-bonding spectral intensities, the bonding-antibonding spectral intensities,
the antibonding-bonding spectral intensities, and the antibonding-antibonding spectral intensities,
respectively. For a better understanding of the characteristic feature of the ARPES autocorrelation in
the bilayer cuprate superconductors, we first plot the intensity map of the components of the ARPES
autocorrelation (a) ${\bar C}^{(11)} ({\bf q},\omega)$, (b) ${\bar C}^{(12)}({\bf q},\omega)$, (c)
${\bar C} ^{(21)} ({\bf q},\omega)$, and (d) ${\bar C}^{(22)}({\bf q},\omega)$ in the $[q_{x},q_{y}]$
plane for the binding-energy $\omega=0.12J$ at $\delta=0.15$ with $T=0.002J$ in
Fig. \ref{autocorrelation-maps}. Obviously, the sharp peaks or so-called discrete spots emerge in
all components ${\bar C}^{(11)}({\bf q},\omega)$, ${\bar C}^{(12)}({\bf q},\omega)$,
${\bar C}^{(21)}({\bf q},\omega)$, and ${\bar C}^{(22)}({\bf q},\omega)$ at almost the same positions,
where the joint density of states is highest. Moreover, these discrete spots in
${\bar C}^{(11)}({\bf q},\omega)$, ${\bar C}^{(12)}({\bf q},\omega)$, ${\bar C}^{(21)}({\bf q},\omega)$,
and ${\bar C}^{(22)}({\bf q},\omega)$ can be described by the {\it octet} scattering model shown in
Fig. \ref{spectral-maps}, and therefore are directly correlated with the wave vectors
${\bf q}^{(BB)}_{i}$ connecting the bonding-bonding hot spots, ${\bf q}^{(BA)}_{i}$ connecting the
bonding-antibonding hot spots, ${\bf q}^{(AB)}_{i}$ connecting the antibonding-bonding hot spots, and
${\bf q}^{(AA)}_{i}$ connecting the antibonding-antibonding hot spots, respectively. In particular,
the autocorrelation pattern in each component ${\bar C}^{(\nu\nu')} ({\bf q},\omega)$ is in a striking
analogy to that in the single-layer case \cite{He14,Gao19}.

\begin{figure}[t!]
\centering
\includegraphics[scale=0.75]{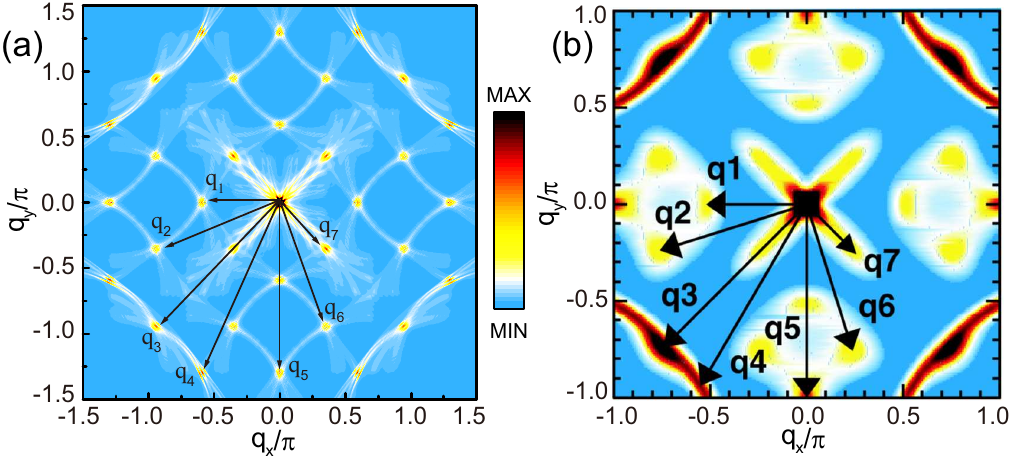}
\caption{(Color online) (a) The intensity map of the ARPES autocorrelation in a $[q_{x},q_{y}]$ plane
for $\omega=0.12J$ and $\delta=0.15$ with $T=0.002J$ for $t/J=2.5$, $t'/t=0.28$, $t_{\perp}/t=0.3$, and
$J=100$ meV. (b) The corresponding experimental results of the optimally doped
Bi$_{2}$Sr$_{2}$CaCu$_{2}$O$_{8+\delta}$ for $\omega=12$ meV taken from Ref. \onlinecite{Chatterjee06}.
\label{autocorrelation-maps-sum}}
\end{figure}

With the above results in Fig. \ref{autocorrelation-maps}, we now turn to discuss the experimentally
measurable ARPES autocorrelation ${\bar C}({\bf q},\omega)$ in the bilayer cuprate superconductors.
In Fig. \ref{autocorrelation-maps-sum}a, we plot the intensity map of the ARPES autocorrelation
${\bar C}({\bf q},\omega)$ in a $[q_{x},q_{y}]$ plane for the binding-energy $\omega=12$ meV at $\delta=0.15$
with $T=0.002J$. For a comparison, the corresponding experimental result \cite{Chatterjee06} of the ARPES
autocorrelation observed from the optimally doped bilayer cuprate superconductor
Bi$_{2}$Sr$_{2}$CaCu$_{2}$O$_{8+\delta}$ for the bind-energy $\omega=12$ meV is also shown in
Fig. \ref{autocorrelation-maps-sum}b. It is very interesting to note that the obtained autocorrelation
pattern in Fig. \ref{autocorrelation-maps-sum}a is well consistent with the corresponding pattern in
Fig. \ref{autocorrelation-maps-sum}b obtained from the ARPES experimental observation on the bilayer
superconductors \cite{Chatterjee06}. Moreover, in comparison with the autocorrelation pattern for each
component ${\bar C}^{(\nu\nu')}({\bf q},\omega)$ in Fig. \ref{autocorrelation-maps}, it is thus shown clearly
that the weight of the ARPES autocorrelation peaks (then the discrete spots) are enhanced, reflecting a fact
that the quasiparticle scattering processes with the scattering wave vectors ${\bf q}_{i}$ shown in
Fig. \ref{spectral-maps} are enhanced, including the enhancement of the SC correlation as we have mentioned
in Sec. \ref{charge-order}. It therefore would be reasonable to expect its connection with the higher
$T_{\rm c}$ in the bilayer cuprate superconductors. In other words, this may be why the optimized $T_{\rm c}$
in the bilayer cuprate superconductors is higher than that in the single-layer case.

\begin{figure}[t!]
\centering
\includegraphics[scale=0.5]{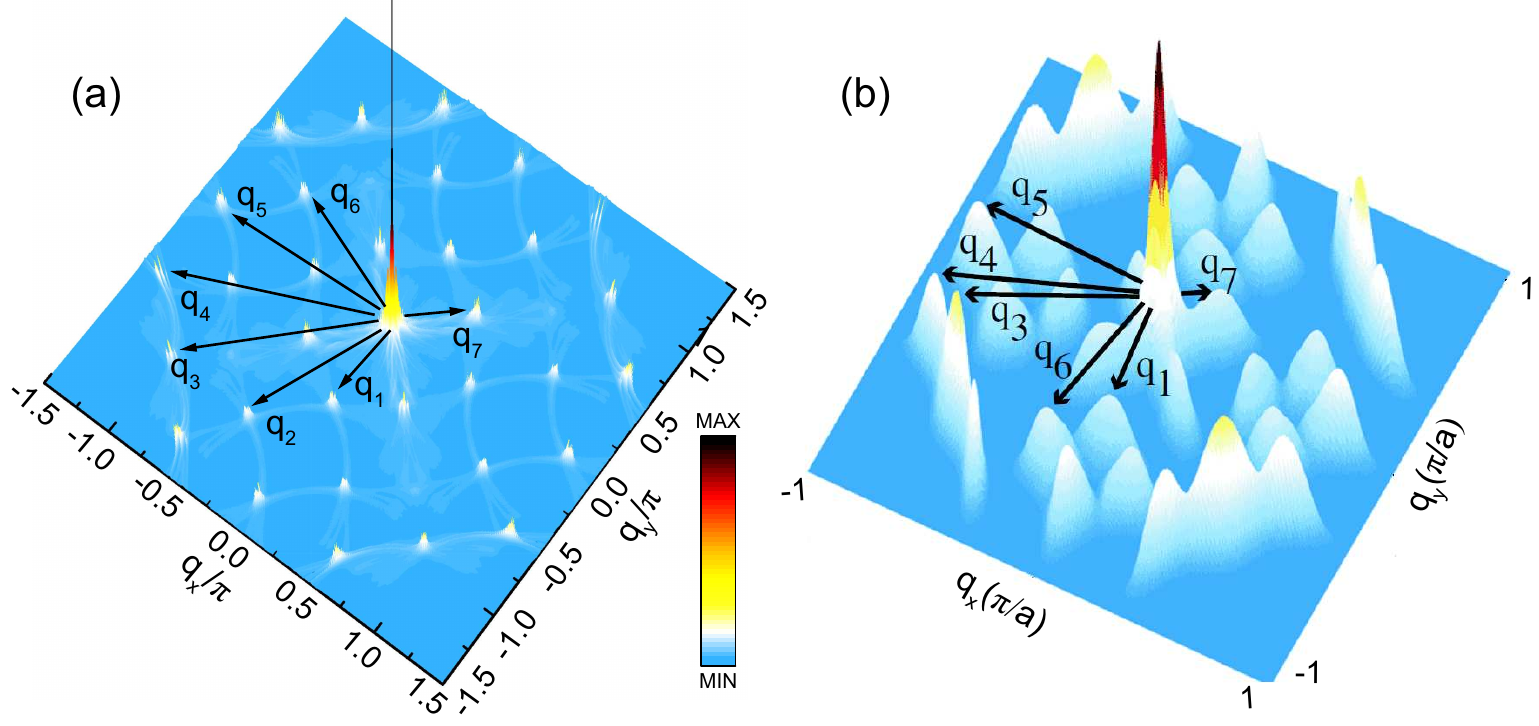}
\caption{(Color online) (a) The intensity map of the ARPES autocorrelation in momentum-space for
$\omega=0.18J$ at $\delta=0.15$ with $T=0.002J$ for $t/J=2.5$, $t'/t=0.28$, $t_{\perp}/t=0.3$, and
$J=100$ meV. (b) The corresponding experimental result of the optimally doped
Bi$_{2}$Sr$_{2}$CaCu$_{2}$O$_{8+\delta}$ for $\omega=18$ meV taken from Ref. \onlinecite{Chatterjee06}.
\label{autocorrelation-peaks}}
\end{figure}

To see these sharp peaks with enhanced weights at discrete spots more clearly, we plot the intensity map
of ${\bar C}({\bf q},\omega)$ in the $[q_{x},q_{y}]$ plane for the binding-energy $\omega=18$ meV at
$\delta=0.15$ with $T=0.002J$ in Fig. \ref{autocorrelation-peaks}a in comparison with the corresponding
experimental result \cite{Chatterjee06} of the ARPES autocorrelation of the optimally doped
Bi$_{2}$Sr$_{2}$CaCu$_{2}$O$_{8+\delta}$ for the binding-energy $\omega=18$ meV in
Fig. \ref{autocorrelation-peaks}b, where the sharp autocorrelation peaks with the enhanced weights are
located exactly at the corresponding discrete spots of ${\bar C}({\bf q},\omega)$, which further confirms
that the sharp peaks in ${\bar C}({\bf q},\omega)$ are closely associated with the corresponding wave vectors
${\bf q}_{i}$ connecting the antibonding and bonding hot spots shown in Fig. \ref{spectral-maps}. To analyze
the evolution of the dispersion of the ARPES autocorrelation peaks, we have also made a series of calculations
for the momentum and energy dependence of ${\bar C}({\bf q},\omega)$ with different ${\bf q}_{i}$, and the
results show that the sharp autocorrelation peaks in ${\bar C}({\bf q},\omega)$ disperse smoothly with energy,
and these dispersion autocorrelation peaks follow the same evolution of the corresponding hot spots with energy,
also in good agreement with the ARPES experimental observations \cite{Chatterjee06}.

\begin{figure}[t!]
\centering
\includegraphics[scale=0.8]{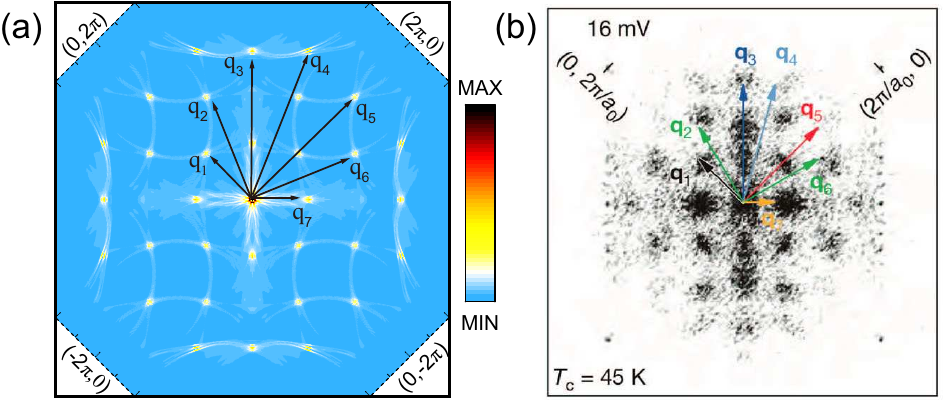}
\caption{(Color online) (a) The intensity map of the ARPES autocorrelation in the $[q_{x},q_{y}]$ plane for
$\omega=0.16J$ at $\delta=0.15$ with $T=0.002J$ for $t/J=2.5$, $t'/t=0.28$, $t_{\perp}/t=0.3$, and $J=100$ meV.
(b) The experimental result of the quasiparticle scattering interference pattern $Z({\bf q},\omega=16$ meV)
for the optimally doped Bi$_{2}$Sr$_{2}$CaCu$_{2}$O$_{8+\delta}$ taken from Ref. \onlinecite{Kohsaka08}.
\label{autocorrelation-octet}}
\end{figure}

Now we turn to discuss the direct connection between the discrete spots in the ARPES autocorrelation
observed from the ARPES measurements and the QSI peaks measured from the FT-STS experiments in the bilayer
cuprtae superconductors. To see this connection clearly, we plot the intensity map of
${\bar C} ({\bf q},\omega)$ in the $[q_{x},q_{y}]$ plane for the binding-energy $\omega=16$ meV at
$\delta=0.15$ with $T=0.002J$ in Fig. \ref{autocorrelation-octet}a. For comparison, the corresponding
experimental result \cite{Kohsaka08} of the QSI pattern obtained from the optimally doped bilayer cuprate
superconductor Bi$_{2}$Sr$_{2}$CaCu$_{2}$O$_{8+\delta}$ for the binding-energy $\omega=16$ meV is also
shown in Fig. \ref{autocorrelation-octet}b. Apparently, the obtained result of the momentum-space structure
of the ARPES autocorrelation pattern ${\bar C}({\bf q}, \omega)$ in Fig. \ref{autocorrelation-octet}a is
well consistent with the corresponding experimental result of the momentum-space structure of the QSI
pattern shown in Fig. \ref{autocorrelation-octet}b. The combined results in Fig. \ref{spectral-maps},
Fig. \ref{autocorrelation-maps-sum}, and Fig. \ref{autocorrelation-octet} therefore indicate that
the {\it octet} scattering model with the scattering wave vectors ${\bf q}_{i}$ connecting the corresponding
hot spots that can give a consistent description of the regions of the highest joint density of states in the
ARPES autocorrelation can be also used to explain the QSI data observed from the FT-STS experiments.
The qualitative agreement between the momentum-space structure of the ARPES autocorrelation pattern and the
momentum-space structure of the QSI pattern therefore confirms an intimate connection between the ARPES
autocorrelation and QSI in the bilayer cuprate superconductors.

The quasiparticles scattering from impurities in cuprate superconductors interfere with one another,
producing a standing wave pattern in the inhomogeneous part $\delta\rho^{(\nu\nu')}({\bf q},\omega)$
for FT LDOS \cite{Pan01,Hoffman02,Kohsaka07,Kohsaka08,Hamidian16}. The dispersion of the peaks in
$\delta\rho^{(\nu\nu')}({\bf q},\omega)$ as a function of bias voltage $\omega$ is analyzed in terms of
the octet scattering model and yields the information about EFS. In the single-layer cuprate superconductors,
the inhomogeneous part $\delta\rho({\bf q},\omega)$ for FT LDOS in the presence of a single point-like
impurity scattering potential has been evaluated based on the kinetic-energy-driven SC mechanism \cite{Gao19},
and then the main features of QSI are qualitatively reproduced. In particular, it has been shown that
in the case of the presence of the strong scattering potential \cite{Gao19}, the momentum-space structure
of the QSI patterns is well consistent with the momentum-space structure of the ARPES autocorrelation patterns.
The essential physics of the intimate connection between the ARPES autocorrelation and QSI in the bilayer
cuprate superconductors is the same as in the single-layer case \cite{Gao19} except for the enhancement of the
weights of the ARPES autocorrelation peaks in the presence of the bilayer coupling. This follows a fact that
although the electron quasiparticle excitation spectrum is split into its bonding and antibonding components
by the bilayer coupling in Eq. (\ref{interlayer-hopping}), the bonding component of the electron quasiparticle
excitation spectrum is independent of the antibonding one. In this case, the essential properties of the
inhomogeneous part $\delta\rho^{(\nu\nu')}({\bf q}, \omega)$ for FT LDOS obtained in terms of the antibonding
and bonding components of the electron normal and anomalous Green's functions (then the antibonding and
bonding components of the electron spectral function) in the bilayer cuprate superconductors in the presence
of a single point-like impurity scattering potential is almost the same as that in the single-layer case in
the presence of a single point-like impurity scattering potential \cite{Gao19}, and then the experimentally
measurable inhomogeneous part $\delta\rho({\bf q},\omega)=\sum_{\nu\nu'}\delta \rho^{(\nu\nu')}({\bf q},\omega)$
for FT LDOS is also almost the same as in the single-layer case, although the weights of the QSI peaks are
enhanced as in the case of the enhanced weights of the ARPES autocorrelation peaks. This is why as in the
single-layer case \cite{Gao19}, the momentum-space structure of the QSI patterns in the bilayer cuprate
superconductors is also well consistent with the momentum-space structure of the ARPES autocorrelation patterns
in the case of the presence of the strong scattering potential. Therefore the intimate connection between the
ARPES autocorrelation and QSI is a universal feature of cuprate superconductors.


\section{Summary and discussions}\label{conclusions}

Within the framework of the kinetic-energy driven superconductivity, we have studied the renormalization
of the electrons in the bilayer cuprate superconductors by taking into account the effect due to the presence
of the bilayer coupling, where the obtained characteristic features can be summarized clearly as the follows:
(a) the electron quasiparticle excitation spectrum in the bilayer cuprate superconductors is split into its
bonding and antibonding components by the bilayer coupling, with each component that is independent, namely,
although two copper-oxide layers within a unit cell are hybridized to form the bonding and antibonding layers,
the bonding component of the electron quasiparticle excitation spectrum in the bonding layer is independent
of the antibonding component of the electron quasiparticle excitation spectrum in the antibonding layer;
(b) in the underdoped and optimally doped regimes, although the bonding (antibonding) EFS contour is truncated
to form the disconnected bonding (antibonding) Fermi arcs, the renormalization from the quasiparticle scattering
further reduces almost all spectral weight in the bonding (antibonding) Fermi arcs to the tips of the bonding
(antibonding) Fermi arcs, which in this case coincide with the bonding (antibonding) hot spots; (c) these
bonding and antibonding hot spots connected by the scattering wave vectors ${\bf q}_{i}$ construct an {\it octet}
scattering model, and then the enhancement of the quasiparticle scattering processes with these scattering wave
vectors ${\bf q}_{i}$ and the SC correlation are confirmed via the result of the ARPES autocorrelation; (d) in a
striking analogy to the single-layer cuprate superconductors, the PDH structure developed in each component of
the electron quasiparticle excitation spectrum along the corresponding EFS is directly related with the peak
structure in the corresponding quasiparticle scattering rate except for at around the hot spots, where the PDH
structure is caused mainly by the pure bilayer coupling, and is in a clear contrast to the single-layer case;
(e) although the kink in the electron quasiparticle dispersion is present all around the antibonding EFS, when
the momentum moves from the node to the antinode, the kink energy smoothly decreases, while the dispersion kink
becomes more pronounced, and in particular, near the cut close to the antinode, develops into a break separating
of the fasting dispersing high-energy part of the electron quasiparticle excitation spectrum from the slower
dispersing low-energy part. By comparing with the corresponding results in the single-layer case, our present
results also indicate that the exotic features of the renormalization of the electrons in the bilayer cuprate
superconductors is particularly obvious due to the presence of the bilayer coupling, reflecting a fact that the
bilayer interaction has significant contributions to the renormalization of the electrons in the bilayer cuprate
superconductors.

\section*{Acknowledgements}

The authors would like to thank Professor Yongjun Wang for helpful discussions. This work was supported
by the National Key Research and Development Program of China under Grant No. 2016YFA0300304, and the
National Natural Science Foundation of China under Grant Nos. 11974051, 11734002, and 11574032.

\vskip 0.5cm

\begin{appendix}

\section{Bonding and antibonding electron self-energies in the bilayer cuprate superconductors}
\label{bonding-antibonding-Green-functions}

In this Appendix, we derive explicitly the bonding and antibonding electron self-energies
$\Sigma_{\rm ph}^{(1)}({\bf k},\omega)$ and $\Sigma_{\rm ph}^{(2)}({\bf k},\omega)$ in the particle-hole
channel and the bonding and antibonding electron self-energies $\Sigma_{\rm pp}^{(1)}({\bf k}, \omega)$
and $\Sigma_{\rm pp}^{(2)}({\bf k},\omega)$ in the particle-particle channel in Eq. (\ref{EGFS}) of the
main text. Following the kinetic-energy driven superconductivity \cite{Feng0306,Feng12,Feng15}, it has
been shown \cite{Lan13} that for the bilayer $t$-$J$ model (\ref{CSS-bilayer-tJ-model}), the interaction
between the charge carriers directly from the kinetic energy by the exchange of spin excitations induces
the charge-carrier pairing state, where the self-consistent equations that satisfied by the charge-carrier
normal and anomalous Green's functions have been obtained as,
\begin{subequations}\label{CCSCES}
\begin{eqnarray}
\tilde{g}({\bf k},\omega)&=&\tilde{g}^{(0)}({\bf k},\omega)+\tilde{g}^{(0)}({\bf k},\omega)
[\tilde{\Sigma}^{(\rm h)}_{\rm ph}({\bf k},\omega)\tilde{g}({\bf k},\omega) \nonumber\\
&-& \tilde{\Sigma}^{(\rm h)}_{\rm pp}({\bf k},\omega)\tilde{\Gamma}^{\dagger}({\bf k},\omega)], ~~~~~\\
\tilde{\Gamma}^{\dagger}({\bf k},\omega)&=&\tilde{g}^{(0)}({\bf k},-\omega)
[\tilde{\Sigma}^{(\rm h)}_{\rm ph}({\bf k},-\omega)\tilde{\Gamma}^{\dagger}({\bf k},\omega)\nonumber\\
&+& \tilde{\Sigma}^{(\rm h)}_{\rm pp}({\bf k},\omega)\tilde{g}({\bf k},\omega)],~~~~
\end{eqnarray}
\end{subequations}
where
$\tilde{g}^{(0)}({\bf k},\omega)=g^{(0)}_{\rm L}({\bf k},\omega)+\sigma_{x}g^{(0)}_{\rm T}({\bf k},\omega)$
is the MF charge-carrier normal Green's function, with the corresponding longitudinal and transverse parts
$g^{(0)}_{\rm L}({\bf k},\omega)$ and $g^{(0)}_{\rm T}({\bf k},\omega)$, respectively, that have been
given explicitly in Ref. \onlinecite{Lan13}, while
$\tilde{\Sigma}^{(\rm h)}_{\rm ph}({\bf k},\omega)=\Sigma^{(\rm h) }_{\rm phL}({\bf k},\omega)
+\Sigma^{(\rm h)}_{\rm phT}({\bf k},\omega)\sigma_{x}$ and
$\tilde{\Sigma}^{(\rm h)}_{\rm pp}({\bf k},\omega)=\Sigma^{(\rm h)}_{\rm ppL}({\bf k},\omega)
+\Sigma^{(\rm h)}_{\rm ppT}({\bf k},\omega)\sigma_{x}$ are the charge-carrier self-energies in the
particle-hole and particle-particle channels, respectively, with $\Sigma^{(\rm h)}_{\rm phL}({\bf k},\omega)$
and $\Sigma^{(\rm h)}_{\rm phT}({\bf k},\omega)$ that are the corresponding longitudinal and transverse parts
of the charge-carrier self-energy in the particle-hole channel, and $\Sigma^{(\rm h)}_{\rm ppL}({\bf k},\omega)$
and $\Sigma^{(\rm h)}_{\rm ppT}({\bf k},\omega)$ that are the corresponding longitudinal and transverse parts
of the charge-carrier self-energy in the particle-particle channel. In particular, these longitudinal and
transverse parts of the charge-carrier self-energies have been obtained explicitly in terms of the spin
bubble as \cite{Lan13},
\begin{widetext}
\begin{subequations}\label{CCSE}
\begin{eqnarray}
\Sigma^{({\rm h})}_{\rm phL}({\bf k},i\omega_{n})&=&{1\over N^{2}}\sum_{{\bf p},{\bf q}}
[R^{(1)}_{{\bf p}+{\bf q}+{\bf k}}{1\over\beta}\sum_{ip_{m}}g_{\rm L} ({{\bf p}+{\bf k}},ip_{m}+i\omega_{n})
\Pi_{\rm LL}({\bf p},{\bf q},ip_{m})\nonumber \\
&+&R^{(2)}_{{\bf p}+{\bf q}+{\bf k}}{1\over\beta}\sum_{ip_{m}}g_{\rm T}({{\bf p}+{\bf k}},ip_{m}+i\omega_{n})
\Pi_{\rm TL}({\bf p},{\bf q},ip_{m})],\\
\Sigma^{({\rm h})}_{\rm phT}({\bf k},i\omega_{n})&=&{1\over N^{2}}\sum_{{\bf p},{\bf q}}
[R^{(1)}_{{\bf p}+{\bf q}+{\bf k}}{1\over\beta}\sum_{ip_{m}} g_{\rm T}({\bf p}+{\bf k},ip_{m}+i\omega_{n})
\Pi_{\rm TT}({\bf p},{\bf q},ip_{m})\nonumber\\
&+&R^{(2)}_{{\bf p}+{\bf q}+{\bf k}}{1\over\beta}\sum_{ip_{m}}g_{\rm L}({\bf p}+{\bf k},ip_{m}+i\omega_{n})
\Pi_{\rm LT}({\bf p},{\bf q},ip_{m})],\\
\Sigma^{({\rm h})}_{\rm ppL}({\bf k},i\omega_{n})&=&{1\over N^{2}}\sum_{{\bf p},{\bf q}}
[R^{(1)}_{{\bf p}+{\bf q}+{\bf k}}{1\over\beta}\sum_{ip_{m}}
\Gamma^{\dagger}_{\rm L}({\bf p}+{\bf k},ip_{m}+i\omega_{n})\Pi_{\rm LL}({\bf p},{\bf q},ip_{m})\nonumber \\
&+& R^{(2)}_{{\bf p}+{\bf q}+{\bf k}}{1\over\beta}\sum_{ip_{m}}
\Gamma^{\dagger}_{\rm T}({\bf p}+{\bf k},ip_{m}+i\omega_{n})\Pi_{\rm TL}({\bf p},{\bf q},ip_{m})],
\end{eqnarray}
\begin{eqnarray}
\Sigma^{({\rm h})}_{\rm ppT}({\bf k},i\omega_{n})&=&{1\over N^{2}}\sum_{{\bf p},{\bf q}}
[R^{(1)}_{{\bf p}+{\bf q}+{\bf k}}{1\over\beta}\sum_{ip_{m}}
\Gamma^{\dagger}_{\rm T}({\bf p}+{\bf k},ip_{m}+i\omega_{n})\Pi_{\rm TT}({\bf p},{\bf q},ip_{m})\nonumber \\
&+&R^{(2)}_{{\bf p}+{\bf q}+{\bf k}}{1\over\beta}\sum_{ip_{m}}
\Gamma^{\dagger}_{\rm L}({\bf p}+{\bf k},ip_{m}+i\omega_{n})\Pi_{\rm LT}({\bf p},{\bf q}, ip_{m})],~~~~~~
\end{eqnarray}
\end{subequations}
\end{widetext}
where $\omega_{n}$ and $p_{m}$ are the fermionic and bosonic Matsubara frequencies, respectively,
$R^{(1)}_{\bf k}=[Z(t\gamma_{\bf k}-t'\gamma'_{\bf k})]^{2}+t_{\perp}^{2}({\bf k})$,
$R^{(2)}_{\bf k}=2Z(t\gamma_{\bf k}-t'\gamma'_{\bf k})t_{\perp}({\bf k})$, and the spin bubbles
$\Pi_{\alpha,\alpha'}({\bf p},{\bf q},ip_{m})$ [$\alpha=L,T$, $\alpha'=L,T$] has been evaluated
explicitly as \cite{Lan13},
\begin{eqnarray}
\Pi_{\alpha\alpha'}({\bf p},{\bf q},ip_{m})&=&(1/\beta)\sum_{iq_{m}}D^{(0)}_{\alpha}({\bf q},iq_{m})\nonumber\\
&\times& D^{(0)}_{\alpha'}({\bf q+p},iq_{m}+ip_{m}),
\end{eqnarray}
with the bosonic Matsubara frequency $q_{m}$, and the corresponding longitudinal and transverse parts of
the spin Green's function $D^{(0)}_{L}({\bf q},\omega)$ and $D^{(0)}_{T}({\bf q},\omega)$ in the self-consistent
renormalized MF level, respectively, that have been evaluated as,
\begin{subequations}
\begin{eqnarray}
D^{(0)}_{L}({\bf k},\omega)&=&{1\over 2}\sum_{\nu=1,2}{B^{(\nu)}_{\bf k}\over\omega^{2}-\omega^{(\nu)2}_{\bf k}}, \\
D^{(0)}_{T}({\bf k},\omega)&=&{1\over 2}\sum_{\nu=1,2}(-1)^{\nu+1}
{B^{(\nu)}_{\bf k}\over\omega^{2}-\omega^{(\nu)2}_{\bf k}},
\end{eqnarray}
\end{subequations}
where the spin excitation spectrum $\omega^{(\nu)}_{\bf k}$ and function $B^{(\nu)}_{\bf k}$ have been given
explicitly in Ref. \onlinecite{Lan13}.

Based on the kinetic-energy-driven SC mechanism, we \cite{Feng15a} have developed a full charge-spin
recombination scheme, where a charge carrier and a localized spin are fully recombined into a constrained electron.
In particular, within this full charge-spin recombination scheme, we have realized that the coupling form between
the electrons and spin excitations is the same as that between the charge carriers and spin excitations, which
therefore implies that the form of the self-consistent equations satisfied by the electron normal and anomalous
Green's functions is the same as the form in Eq. (\ref{CCSCES}) satisfied by the charge-carrier normal and anomalous
Green's functions. In other words, we can perform a full charge-spin recombination in which the charge-carrier normal
and anomalous Green's functions $\tilde{g}({\bf k},\omega)$ and $\tilde{\Gamma}^{\dagger}({\bf k}, \omega)$ in
Eq. (\ref{CCSCES}) are replaced by the electron normal and anomalous Green's functions $\tilde{G}({\bf k},\omega)$
and $\tilde{\Im}^{\dagger}({\bf k},\omega)$, respectively, and then we obtained explicitly the self-consistent
equations satisfied by the electron normal and anomalous Green's functions of the bilayer $t$-$J$
model (\ref{bilayer-tJ-model}) as,
\begin{subequations}\label{ESCES}
\begin{eqnarray}
\tilde{G}({\bf k},\omega)&=&\tilde{G}^{(0)}({\bf k},\omega)+\tilde{G}^{(0)}({\bf k},\omega)
[\tilde{\Sigma}_{\rm ph}({\bf k},\omega)\tilde{G}({\bf k}, \omega) \nonumber\\
&-&\tilde{\Sigma}_{\rm pp}({\bf k},\omega)\tilde{\Im}^{\dagger}({\bf k},\omega)], ~~~~~\label{EDGF} \\
\tilde{\Im}^{\dagger}({\bf k},\omega)&=&\tilde{G}^{(0)}({\bf k},-\omega)
[\tilde{\Sigma}_{\rm ph}({\bf k},-\omega)\tilde{\Im}^{\dagger}({\bf k}, \omega) \nonumber\\
&+&\tilde{\Sigma}_{\rm pp}({\bf k},\omega)\tilde{G}({\bf k}, \omega)], ~~~~~\label{EODGF}
\end{eqnarray}
\end{subequations}
where $\tilde{G}^{(0)}({\bf k},\omega)=G^{(0)}_{\rm L}({\bf k},\omega)+\sigma_{x}G^{(0)}_{\rm T}({\bf k},\omega)$
is the MF electron normal Green's function, with the corresponding longitudinal and transverse parts
$G^{(0)}_{\rm L}({\bf k},\omega)$ and $G^{(0)}_{\rm T}({\bf k},\omega)$, respectively, that can be obtained
directly from the bilayer $t$-$J$ model (\ref{bilayer-tJ-model}) as,
\begin{subequations}\label{MFEGFS}
\begin{eqnarray}
G_{\rm L}^{(0)}({\bf k},\omega)&=&{1\over 2}\sum_{\nu=1,2}{1\over\omega-\varepsilon^{(\nu)}_{\bf k}}, \\
G_{\rm T}^{(0)}({\bf k},\omega)&=&{1\over 2}\sum_{\nu=1,2}(-1)^{\nu+1}{1\over\omega-\varepsilon^{(\nu)}_{\bf k}},
\end{eqnarray}
\end{subequations}
while $\tilde{\Sigma}_{\rm ph}({\bf k},\omega)=\Sigma_{\rm phL}({\bf k},\omega)
+\sigma_{x}\Sigma_{\rm phT}({\bf k},\omega)$ and
$\tilde{\Sigma}_{\rm pp}({\bf k},\omega)=\Sigma_{\rm ppL}({\bf k},\omega)
+\sigma_{x}\Sigma_{\rm ppT}({\bf k},\omega)$ are electron self-energies in the particle-hole and
particle-particle channels, respectively, with the corresponding longitudinal and transverse parts of the
electron self-energies $\Sigma_{\rm phL} ({\bf k},\omega)$ [$\Sigma_{\rm ppL} ({\bf k},\omega)$] and
$\Sigma_{\rm phT}({\bf k},\omega)$ [$\Sigma_{\rm ppT} ({\bf k},\omega)$], respectively, in the particle-hole
(particle-particle) channel are obtained directly from the corresponding parts of the charge-carrier
self-energies in Eq. (\ref{CCSE}) by the replacement of the charge-carrier normal and anomalous Green's
functions $\tilde{g}({\bf k},\omega)$ and $\tilde{\Gamma}^{\dagger}({\bf k},\omega)$ with the corresponding
electron normal and anomalous Green's functions $\tilde{G}({\bf k},\omega)$ and
$\tilde{\Im}^{\dagger}({\bf k},\omega)$ as,
\begin{widetext}
\begin{subequations}\label{ESE-LT}
\begin{eqnarray}
\Sigma_{\rm phL}({\bf k},i\omega_{n})&=&{1\over N^{2}}\sum_{{\bf p},{\bf q}}
[R^{(1)}_{{\bf p}+{\bf q}+{\bf k}}{1\over\beta}\sum_{ip_{m}}G_{\rm L} ({{\bf p}+{\bf k}},ip_{m}+i\omega_{n})
\Pi_{\rm LL}({\bf p},{\bf q},ip_{m})\nonumber \\
&+&R^{(2)}_{{\bf p}+{\bf q}+{\bf k}}{1\over\beta}\sum_{ip_{m}}G_{\rm T}({{\bf p}+{\bf k}},ip_{m}+i\omega_{n})
\Pi_{\rm TL}({\bf p},{\bf q},ip_{m})],
\end{eqnarray}
\begin{eqnarray}
\Sigma_{\rm phT}({\bf k},i\omega_{n})&=&{1\over N^{2}}\sum_{{\bf p},{\bf q}}
[R^{(1)}_{{\bf p}+{\bf q}+{\bf k}}{1\over\beta}\sum_{ip_{m}}G_{\rm T} ({\bf p}+{\bf k},ip_{m}+i\omega_{n})
\Pi_{\rm TT}({\bf p},{\bf q},ip_{m})\nonumber\\
&+&R^{(2)}_{{\bf p}+{\bf q}+{\bf k}}{1\over\beta}\sum_{ip_{m}}G_{\rm L}({\bf p}+{\bf k},ip_{m}+i\omega_{n})
\Pi_{\rm LT}({\bf p},{\bf q},ip_{m})],\\
\Sigma_{\rm ppL}({\bf k},i\omega_{n})&=&{1\over N^{2}}\sum_{{\bf p},{\bf q}}
[R^{(1)}_{{\bf p}+{\bf q}+{\bf k}}{1\over \beta}\sum_{ip_{m}}
\Im^{\dagger}_{\rm L}({\bf p}+{\bf k},ip_{m}+i\omega_{n})\Pi_{\rm LL}({\bf p},{\bf q},ip_{m})\nonumber \\
&+& R^{(2)}_{{\bf p}+{\bf q}+{\bf k}}{1\over\beta}\sum_{ip_{m}}
\Im^{\dagger}_{\rm T}({\bf p}+{\bf k},ip_{m}+i\omega_{n})\Pi_{\rm TL}({\bf p},{\bf q}, ip_{m})],\\
\Sigma_{\rm ppT}({\bf k},i\omega_{n})&=&{1\over N^{2}}\sum_{{\bf p},{\bf q}}
[R^{(1)}_{{\bf p}+{\bf q}+{\bf k}}{1\over\beta}\sum_{ip_{m}}
\Im^{\dagger}_{\rm T}({\bf p}+{\bf k},ip_{m}+i\omega_{n})\Pi_{\rm TT}({\bf p},{\bf q},ip_{m})\nonumber \\
&+&R^{(2)}_{{\bf p}+{\bf q}+{\bf k}}{1\over\beta}\sum_{ip_{m}}
\Im^{\dagger}_{\rm L}({\bf p}+{\bf k},ip_{m}+i\omega_{n})\Pi_{\rm LT}({\bf p},{\bf q}, ip_{m})].~~~~~~
\end{eqnarray}
\end{subequations}
\end{widetext}

However, as we have mentioned in Eq. (\ref{QPE-spectrum}), in the case of the coherent coupling of the
copper-oxide layers within a unit cell, the more appropriate classification in the bilayer coupling case
is in terms of the physical quantities within the basis of the bonding and antibonding components
\cite{Eschrig06}. With the help of the above longitudinal and transverse parts of the electron
self-energies in the particle-hole and particle-particle channels, the corresponding bonding and antibonding
parts of the electron self-energies in the particle-hole and particle-particle channels can be expressed
explicitly in the bonding-antibonding representation as,
\begin{subequations}\label{ESE-BA}
\begin{eqnarray}
\Sigma^{(\nu)}_{\rm ph}({\bf k},\omega)&=&\Sigma_{\rm phL}({\bf k},\omega)+(-1)^{\nu+1}
\Sigma_{\rm phT}({\bf k},\omega),\\
\Sigma^{(\nu)}_{\rm pp}({\bf k},\omega)&=&\Sigma_{\rm ppL}({\bf k},\omega)+(-1)^{\nu+1}
\Sigma_{\rm ppT}({\bf k},\omega),~~~~~~
\end{eqnarray}
\end{subequations}
respectively, and then the corresponding bonding and antibonding parts of the electron normal and anomalous
Green's functions now can be obtained as quoted in Eq. (\ref{EGFS}).

Now our goal is to evaluate these bonding and antibonding parts of the electron self-energies in the
particle-hole and particle-particle channels. According to our previous discussions of the renormalization
of the electrons in the single-layer cuprate superconductors \cite{Feng15a}, the bonding and antibonding
electron self-energies $\Sigma^{(1)}_{\rm pp}({\bf k},\omega)$ and $\Sigma^{(2)}_{\rm pp}({\bf k},\omega)$
in the particle-particle channel represent the bonding and antibonding electron pair gaps, respectively,
while the bonding and antibonding electron self-energies $\Sigma^{(1)}_{\rm ph}({\bf k},\omega)$ and
$\Sigma^{(2)}_{\rm ph}({\bf k},\omega)$ in the particle-hole channel represent the bonding and antibonding
electron quasiparticle coherences, respectively. In particular, the electron self-energy
$\Sigma^{(\nu)}_{\rm pp}({\bf k},\omega)$ is an even function of energy $\omega$, while the electron
self-energy $\Sigma^{(\nu)}_{\rm ph} ({\bf k},\omega)$ is not. For a convenience in the following
calculations, the electron self-energy $\Sigma^{(\nu)}_{\rm ph}({\bf k},\omega)$ can be separated into
two parts as
$\Sigma^{(\nu)}_{\rm ph}({\bf k},\omega)=\Sigma^{(\nu)}_{\rm phe}({\bf k},\omega)
+\omega\Sigma^{(\nu)}_{\rm pho}({\bf k},\omega)$ with the corresponding symmetric part
$\Sigma^{(\nu)}_{\rm phe}({\bf k},\omega)$ and antisymmetric part
$\omega\Sigma^{(\nu)}_{\rm pho}({\bf k},\omega)$, respectively, and then both
$\Sigma^{(\nu)}_{\rm pho}({\bf k},\omega)$ and $\Sigma^{(\nu)}_{\rm phe}({\bf k},\omega)$ are an even
function of energy $\omega$. Following the common practice \cite{Mahan81}, the electron quasiparticle
coherent weight in the bilayer cuprate superconductors therefore can be obtained as
$Z_{\rm F}^{(\nu)-1} ({\bf k},\omega)=Z^{-1}_{\rm F1}({\bf k},\omega)-(-1)^{\nu+1}Z^{-1}_{\rm F2}({\bf k},\omega)$,
with $Z^{-1}_{\rm F1}({\bf k}, \omega)=1-\Sigma_{\rm phLo}({\bf k},\omega)$ and
$Z^{-1}_{\rm F2} ({\bf k},\omega)=\Sigma_{\rm phTo}({\bf k},\omega)$, where
$\Sigma_{\rm phLo}({\bf k},\omega)$ and $\Sigma_{\rm phTo}({\bf k},\omega)$ are the corresponding antisymmetric
parts of the longitudinal and transverse electron self-energies $\Sigma_{\rm phL}({\bf k},\omega)$ and
$\Sigma_{\rm phT} ({\bf k},\omega)$, respectively. In this paper, we mainly discuss the low-energy behavior
of the renormalized electrons in bilayer cuprate superconductors, and therefore the electron pair gap can be
discussed in the static limit, i.e.,
$\bar{\Delta}^{(\nu)}({\bf k})=\Sigma^{(\nu)}_{\rm pp}({\bf k},0)=\bar{\Delta}_{\rm L}({\bf k})
+(-1)^{\nu+1} \bar{\Delta}_{\rm T} ({\bf k})$,
with $\bar{\Delta}_{\rm L}({\bf k})=\Sigma_{\rm ppL}({\bf k},0)=\bar{\Delta}_{\rm L}\gamma^{({\rm d})}_{\bf k}$,
$\gamma^{\rm (d)}_{\bf k}=[\cos k_{x}-\cos k_{y}]/2$, and
$\bar{\Delta}_{\rm T}({\bf k})=\Sigma_{\rm ppT}({\bf k},0)=\bar{\Delta}_{\rm T}$, and the electron
quasiparticle coherent weight
$Z_{\rm F}^{(\nu)-1}({\bf k} ,0)=Z^{-1}_{\rm F1}({\bf k},0)-(-1)^{\nu+1}Z^{-1}_{\rm F2}({\bf k},0)$. Although
the electron quasiparticle coherent weight $Z^{(\nu)}_{\rm F}({\bf k})$ still is a function of ${\bf k}$,
we can follow the ARPES experiments  \cite{DLFeng00,Ding01}, and choice the wave vector ${\bf k}$ in
$Z^{(\nu)}_{\rm F}({\bf k})$ as $Z^{(\nu)}_{\rm F}({\bf k}) \mid_{{\bf k}=[\pi,0]}$.

With the help of the above discussions, the bonding and antibonding parts of the self-consistent renormalized
MF electron normal and anomalous Green's functions now can be obtained directly from Eq. (\ref{EGFS}) as,
\begin{subequations}\label{MF-EGFS}
\begin{eqnarray}
G^{\rm (RMF)}_{\nu}({\bf k},\omega)&=&Z_{\rm F}^{(\nu)}\left ({U^{(\nu)2}_{\bf k}\over\omega
-E^{(\nu)}_{\bf k}}+{V^{(\nu)2}_{\bf k}\over\omega +E^{(\nu)}_{\bf k}} \right ), ~~~~~~~\\
\Im^{{\rm (RMF)}\dagger}_{\nu}({\bf k},\omega)&=&-Z_{\rm F}^{(\nu)}{\bar{\Delta}^{(\nu)}_{\rm Z}({\bf k})
\over 2E^{(\nu)}_{\bf k}}\nonumber\\
&\times& \left ({1\over\omega -E^{(\nu)}_{\bf k}}-{1\over\omega+E^{(\nu)}_{\bf k}}\right ),~~~~~~~~~
\end{eqnarray}
\end{subequations}
where the coherence factors $U^{(\nu)2}_{\bf k}=(1+\bar{\varepsilon}^{(\nu)}_{\bf k}/E^{(\nu)}_{\bf k})/2$
and $V^{(\nu)2}_{\bf k}=(1- \bar{\varepsilon}^{(\nu)}_{\bf k}/E^{(\nu)}_{\bf k})/2$,
$\bar{\varepsilon}^{(\nu)}_{\bf k}=Z_{\rm F}^{(\nu)}\varepsilon^{(\nu)}_{\bf k}$,
$E^{(\nu)}_{\bf k}=\sqrt {[\bar{\varepsilon}^{(\nu)}_{\bf k}]^{2}+[\bar{\Delta}^{(\nu)}_{\rm Z}({\bf k})]^{2}}$,
and $\bar{\Delta}^{(\nu)}_{\rm Z} ({\bf k}) =Z_{\rm F}^{(\nu)} \bar{\Delta}^{(\nu)}({\bf k})$. In this case,
the bare dispersion relation $\varepsilon^{(\nu)}_{\bf k}$ has been renormalized into
$\bar{\varepsilon}^{(\nu)}_{\bf k}$ by the electron quasiparticle coherent weight $Z^{(\nu)}_{\rm F}$,
which reduces the electron quasiparticle bandwidth, and then the energy scale of the electron quasiparticle
band is controlled by the magnetic interaction $J$. In particular, in the normal-state, the electron pair gap
$\bar{\Delta}^{(\nu)}({\bf k})=0$, and then the bonding and antibonding parts of the self-consistent
renormalized MF electron normal Green's functions in Eq. (\ref{MF-EGFS}) are reduced as,
\begin{eqnarray}\label{Normal-MF-EGFS}
G^{\rm (RMF)}_{\nu}({\bf k},\omega)&=&Z_{\rm F}^{(\nu)}{1\over\omega-\bar{\varepsilon}^{(\nu)}_{\bf k} },
\end{eqnarray}
with the related bonding and antibonding components of the self-consistent renormalized MF electron
quasiparticle excitation spectrum,
\begin{subequations}\label{Normal-MF-ESF}
\begin{eqnarray}
I^{(0)}_{1}({\bf k},\omega)&=&2\pi n_{\rm F}(\omega)Z^{(1)}_{\rm F}\delta(\omega-\bar{\varepsilon}^{(1)}_{\bf k}), \\
I^{(0)}_{2}({\bf k},\omega)&=&2\pi n_{\rm F}(\omega)Z^{(2)}_{\rm F}\delta(\omega-\bar{\varepsilon}^{(2)}_{\bf k}),
\end{eqnarray}
\end{subequations}
respectively.

Substituting these bonding and antibonding parts of the self-consistent renormalized MF electron normal
and anomalous Green's functions in Eq. (\ref{MF-EGFS}) into Eqs. (\ref{ESE-LT}) and (\ref{ESE-BA}), we
therefore obtain explicitly the bonding and antibonding parts of the electron self-energies in the particle-hole
and particle-particle channels as,
\begin{widetext}
\begin{subequations}\label{ESE}
\begin{eqnarray}
\Sigma_{\rm ph}^{(\nu)}({\bf k},\omega)&=&{1\over N^{2}}\sum_{\substack{{\bf pq}\\ \mu\nu_{1}\nu_{2}\nu_{3}}}(-1)^{\mu}
\Omega^{\nu\nu_{1}\nu_{2}\nu_{3}}_{{\bf k}{\bf p}{\bf q}}\left [U_{{\bf k}+{\bf p}}^{(\nu_{1})2}
\left ({F_{\mu\nu_{1}\nu_{2}\nu_{3}}^{(1)}({\bf k}, {\bf p},{\bf q})\over\omega-E_{{\bf k}+{\bf p}}^{(\nu_{1})}
-\omega_{\mu{\bf pq}}^{(\nu_{2}\nu_{3})}}+{F_{\mu\nu_{1}\nu_{2}\nu_{3}}^{(2)}({\bf k}, {\bf p},{\bf q})\over\omega
-E_{{\bf k}+{\bf p}}^{(\nu_{1})}+\omega_{\mu{\bf pq}}^{(\nu_{2}\nu_{3})}}\right )\right . \nonumber\\
&+&\left . V_{{\bf k}+{\bf p}}^{(\nu_{1})2}\left ({F_{\mu\nu_{1}\nu_{2}\nu_{3}}^{(1)}({\bf k},{\bf p},{\bf q})\over\omega
+E_{{\bf k}+{\bf p}} ^{(\nu_{1})}+\omega_{\mu{\bf pq}}^{(\nu_{2}\nu_{3})}}
+{F_{\mu\nu_{1}\nu_{2}\nu_{3}}^{(2)}({\bf k},{\bf p},{\bf q})
\over\omega+E_{{\bf k}+{\bf p}} ^{(\nu_{1})}-\omega_{\mu{\bf pq}}^{(\nu_{2}\nu_{3})}}\right )\right ],
\label{bonding-ESE}\\
\Sigma_{\rm pp}^{(\nu)}({\bf k},\omega)&=&-{1\over N^{2}}\sum_{\substack{{\bf pq}\\ \mu\nu_{1}\nu_{2}\nu_{3}}}(-1)^{\mu}
\Omega^{\nu\nu_{1}\nu_{2}\nu_{3}}_{{\bf k}{\bf p}{\bf q}}{\bar{\Delta}_{\rm Z}^{(\nu_{1})}({\bf k}+{\bf p})
\over 2E^{(\nu_{1})}_{{\bf k}+{\bf p}}}
\left [ \left ({F_{\mu\nu_{1}\nu_{2}\nu_{3}}^{(1)}({\bf k},{\bf p},{\bf q})\over
\omega-E_{{\bf k}+{\bf p}}^{(\nu_{1})}-\omega_{\mu{\bf pq}} ^{(\nu_{2}\nu_{3})}}
+{F_{\mu\nu_{1}\nu_{2}\nu_{3}}^{(2)}({\bf k},{\bf p},{\bf q})\over \omega-E_{{\bf k}+{\bf p}}^{(\nu_{1})}+\omega_{\mu{\bf pq}} ^{(\nu_{2}\nu_{3})}}\right )\right .\nonumber\\
&-&\left . \left ({F_{\mu\nu_{1}\nu_{2}\nu_{3}}^{(1)}({\bf k},{\bf p},{\bf q})\over\omega
+E_{{\bf k}+{\bf p}}^{(\nu_{1})}+\omega_{\mu{\bf pq}} ^{(\nu_{2}\nu_{3})}}
+{F_{\mu\nu_{1}\nu_{2}\nu_{3}}^{(2)}({\bf k},{\bf p},{\bf q})\over\omega
+E_{{\bf k}+{\bf p}}^{(\nu_{1})}-\omega_{\mu{\bf pq}} ^{(\nu_{2}\nu_{3})}}\right )\right ],
\label{antibonding-ESE}
\end{eqnarray}
\end{subequations}
\end{widetext}
respectively, where $\mu=1,2$ and $\nu_{i}=1,2$ with $i=1,2,3$,
$\Omega^{\nu\nu_{1}\nu_{2}\nu_{3}}_{{\bf k}{\bf p}{\bf q}}=\Lambda_{{\bf k}+{\bf p}+{\bf q}}^{\nu\nu_{1}\nu_{2}\nu_{3}}
Z_{\rm F}^{(\nu_{1})}B_{\bf q}^{(\nu_{2})}B_{{\bf p}+{\bf q}}^{(\nu_{3})}
/[32\omega_{\bf q}^{(\nu_{2})}\omega_{{\bf p}+{\bf q}}^{(\nu_{3})}]$,
$\Lambda_{\bf k}^{\nu\nu_{1}\nu_{2}\nu_{3}}=[1+(-1)^{\nu+\nu_{1}+\nu_{2}+\nu_{3}}]
[Z(t\gamma_{\bf k}- t' \gamma'_{\bf k})+(-1)^{\nu+\nu_{3}}t_{\perp}({\bf k})]^{2}$,
$\omega_{\mu{\bf pq}}^{(\nu_{2}\nu_{3})}=\omega_{\bf q}^{(\nu_{2})}-(-1)^{\mu}
\omega_{{\bf p}+{\bf q}}^{(\nu_{3})}$, and the functions,
\begin{eqnarray*}
F_{\mu\nu_{1}\nu_{2}\nu_{3}}^{(1)}({\bf k},{\bf p},{\bf q})&=&n_{\rm F}(E_{{\bf k}+{\bf p}}^{(\nu_{1})})
n^{\mu\nu_{2}\nu_{3}}_{\rm 1B}({\bf p},{\bf q}) +n^{\mu\nu_{2}\nu_{3}}_{\rm 2B}({\bf p},{\bf q}),
\end{eqnarray*}
\begin{eqnarray*}
F_{\mu\nu_{1}\nu_{2}\nu_{3}}^{(2)}({\bf k},{\bf p},{\bf q})&=&[1-n_{\rm F}(E_{{\bf k}+{\bf p}}^{(\nu_{1})})]
n^{\mu\nu_{2}\nu_{3}}_{\rm 1B}({\bf p}, {\bf q})\nonumber\\
&+&n^{\mu\nu_{2}\nu_{3}}_{\rm 2B}({\bf p},{\bf q}),
\end{eqnarray*}
with $n^{\mu\nu_{2}\nu_{3}}_{\rm 1B}({\bf p},{\bf q})=n_{\rm B}(\omega_{\bf q}^{(\nu_{2}) })
-n_{\rm B}[(-1)^{\mu}\omega_{{\bf p}+{\bf q}}^{(\nu_{3})}]$, $n^{\mu\nu_{2}\nu_{3}}_{\rm 2B}({\bf p},{\bf q})
=[1+n_{\rm B}(\omega_{\bf q}^{ (\nu_{2})})]n_{\rm B}[(-1)^{\mu}\omega_{{\bf p}+{\bf q}}^{(\nu_{3})}]$, and the
boson and fermion distribution functions $n_{\rm B}(\omega)$ and $n_{\rm F}(E)$, respectively. In this case,
the longitudinal and transverse parts of the electron pair gap parameter, the electron quasiparticle coherent
weight, and the chemical potential satisfy following self-consistent equations,
\begin{widetext}
\begin{subequations}
\begin{eqnarray}
\bar{\Delta}_{\rm L}&=&{4\over N^{3}}\sum_{\substack{{\bf kpq}\mu\nu\\ \nu_{1}\nu_{2}\nu_{3}}}(-1)^{\mu}
\Omega^{\nu\nu_{1}\nu_{2}\nu_{3}}_{\bf kpq} {\bar{\Delta}_{\rm Z}^{(\nu_{1})}({\bf k}+{\bf p})
\gamma_{\bf k}^{\rm (d)}\over 2E^{(\nu_{1})}_{{\bf k}+{\bf p}}}\left ( {F_{\mu\nu_{1}\nu_{2}\nu_{3}}^{(1)}({\bf k},{\bf p},{\bf q})
\over E_{{\bf k}+{\bf p}}^{(\nu_{1})}+\omega_{\mu{\bf pq}}^{(\nu_{2}\nu_{3})}}
+{F_{\mu\nu_{1}\nu_{2}\nu_{3}}^{(2)}({\bf k},{\bf p},{\bf q})\over E_{{\bf k}+{\bf p}}^{(\nu_{1})}
-\omega_{\mu{\bf pq}}^{(\nu_{2}\nu_{3})}} \right ),
\end{eqnarray}
\begin{eqnarray}
\bar{\Delta}_{\rm T}&=&-{1\over N^{3}}\sum_{\substack{{\bf kpq}\mu\nu\\ \nu_{1}\nu_{2}\nu_{3}}}(-1)^{\mu+\nu}
\Omega^{\nu\nu_{1}\nu_{2}\nu_{3}}_{\bf kpq}{\bar{\Delta}_{\rm Z}^{(\nu_{1})}({\bf k}+{\bf p})\over 2E^{(\nu_{1})}_{{\bf k}+{\bf p}}}
\left ({F_{\mu\nu_{1}\nu_{2}\nu_{3}}^{(1)}({\bf k},{\bf p}, {\bf q})\over
E_{{\bf k}+{\bf p}}^{(\nu_{1})}+\omega_{\mu{\bf pq}}^{(\nu_{2}\nu_{3})}}+{F_{\mu\nu_{1}\nu_{2}\nu_{3}}^{(2)}({\bf k},{\bf p},{\bf q})
\over E_{{\bf k}+{\bf p}}^{(\nu_{1})}-\omega_{\mu{\bf pq}}^{(\nu_{2}\nu_{3})}}\right ),\\
{1\over Z_{\rm F}^{(\nu)}}&=&1+{1\over N^{2}}\sum_{\substack{{\bf pq}\mu\\ \nu_{1}\nu_{2}\nu_{3}}}(-1)^{\mu}
\Omega^{\nu\nu_{1}\nu_{2}\nu_{3}}_{\bf k_{0}pq}\left ( {F_{\mu\nu_{1}\nu_{2}\nu_{3}}^{(1)}({\bf k_{0}},{\bf p},{\bf q})
\over [E_{{\bf k_{0}}+{\bf p}}^{(\nu_{1})}+\omega_{\mu{\bf pq}} ^{(\nu_{2}\nu_{3})}]^{2}}
+{F_{\mu\nu_{1}\nu_{2}\nu_{3}}^{(2)}({\bf k_{0}},{\bf p},{\bf q})
\over [E_{{\bf k_{0}}+{\bf p}}^{(\nu_{1})}-\omega_{\mu{\bf pq}}^{(\nu_{2}\nu_{3})}]^{2}}\right ),\\
1-\delta &=& {1\over 2N}\sum_{{\bf k},\nu}Z_{\rm F}^{(\nu)}\left (1-{\bar{\varepsilon}^{(\nu)}_{\bf k}
\over E^{(\nu)}_{\bf k}}{\rm tanh}[{1\over 2} \beta E^{(\nu)}_{\bf k}] \right ),\label{SCE-3}
\end{eqnarray}
\end{subequations}
\end{widetext}
with ${\bf k}_0=[\pi,0]$. These equations have been solved self-consistently, and then all order parameters
and chemical potential are determined by the self-consistent calculation. In particular, although the shape
of the bonding and antibonding EFS contours determined by the bare band structure
$\varepsilon^{(\nu)}_{\bf k}$ has been changed by the self-consistent renormalized MF band structure
$\bar{\varepsilon}^{(\nu)}_{\bf k}$, the self-consistent equation in Eq. (\ref{SCE-3}) guarantees that
the bonding and antibonding EFS contours determined by the self-consistent renormalized MF band structure
$\bar{\varepsilon}^{(\nu)}_{\bf k}$ satisfy Luttinger's theorem \cite{Luttinger60}, i.e., the effective
area of the bonding and antibonding EFS contours contains $1-\delta$ electrons.

\end{appendix}

\end{document}